\documentclass[a4paper,fleqn,usenatbib]{mn2e}
\usepackage{times}
\long\def\symbolfootnote[#1]#2{\begingroup%
\def\thefootnote{\fnsymbol{footnote}}\footnote[#1]{#2}\endgroup}
\usepackage[T1]{fontenc}
\usepackage{ae,aecompl}

\usepackage{graphicx}	
\usepackage{amsmath}	
\usepackage{amssymb}	
\usepackage{multirow}
\usepackage{natbib}
\defcitealias{kallivayalil13}{K13}
\defcitealias{bechtol15b}{DES15}

\title[A Magellanic Origin of the DES Dwarfs]{A Magellanic Origin of the DES Dwarfs}

\author[P. Jethwa et al.]{
P. Jethwa,$^{1}$\thanks{E-mail: pj253@ast.cam.ac.uk}
D. Erkal$^{1}$\thanks{E-mail: derkal@ast.cam.ac.uk} \& V. Belokurov$^{1}$\thanks{E-mail: vasily@ast.cam.ac.uk}
\\
$^{1}$Institute of Astronomy, University of Cambridge, Madingley Road, Cambridge CB3 0HA, UK\\
}

\date{Accepted XXX. Received YYY; in original form ZZZ}

\pubyear{2015}

\begin{document}
\label{firstpage}
\pagerange{\pageref{firstpage}--\pageref{lastpage}}
\maketitle

\begin{abstract}
We establish the connection between the Magellanic Clouds (MCs) and
the dwarf galaxy candidates discovered in the Dark Energy Survey (DES) by
building a dynamical model of the MC satellite populations, based on
an extensive suite of tailor-made numerical simulations. Our model
takes into account the response of the Galaxy to the MCs infall, the
dynamical friction experienced by the MCs and the disruption of the MC
satellites by their hosts. The simulation suite samples over the
uncertainties in the MC's proper motions, the masses of the MW and the
Clouds themselves and allows for flexibility in the intrinsic volume
density distribution of the MC satellites. As a result, we can
accurately reproduce the DES satellites' observed positions and
kinematics.
Assuming that Milky Way (MW) dwarfs follow the distribution of subhaloes in $\Lambda$CDM, 
we further demonstrate that, of 14 observed satellites, the MW halo
contributes fewer than 4 (8) of these with 68\% (95\%) confidence and that
7 (12) DES dwarfs have probabilities greater than
0.7 (0.5) of belonging to the LMC. Marginalising over the entire suite,
we constrain the total number of the Magellanic satellites at
$\sim70$, the mass of the LMC around $10^{11}$M$_{\odot}$ and show
that the Clouds have likely endured only one Galactic pericentric
passage so far. Finally, we give predictions for the line-of-sight
velocities and the proper motions of the satellites discovered in the
vicinity of the LMC.

\end{abstract}

\begin{keywords}
Galaxies -- galaxies: Magellanic Clouds -- galaxies: dwarf
\end{keywords}


\section{Introduction}

Before clues to the nature of the Dark Matter (DM) and the details of
the power spectrum of density fluctuations in the early Universe were
uncovered, it had already been shown that the small ``seed'' objects
must have collapsed first \citep[][]{peebles65} to later spawn
the assembly of the larger ones. From the bottom up, gravity
fabricates objects that appear self-similar across a large range of
mass scales \citep[][]{press74} - this is a common feature of
an entire class of structure formation scenarios in an expanding
Universe. For the Cold Dark Matter model in particular, the mass
function of bound DM condensations is demonstrated to extend as far
down as the numerical resolution allows; however, the levels of
sub-structure within sub-structure - so called sub-sub-haloes - appear
slightly depleted \citep[see e.g.][]{springel08}. Nonetheless, even if
satellite galaxies are not simply scaled down versions of their hosts,
$\Lambda$CDM stipulates that the most massive of the Milky Way (MW)
dwarfs ought to have dark and indeed luminous \citep[e.g.][]{wheeler15} satellites of their own.

The Large Magellanic Cloud is the most massive satellite of the Milky
Way and, hence, is an obvious place to probe the extension of the
Cosmological hierarchy to the smallest galaxy scales. Last year, the
Dark Energy Survey (DES) provided nineteen new entries to the
inventory of Galactic satellites
\citep{koposov15a,bechtol15a,bechtol15b,kim15a,kim15b,luque15}.  Given
the proximity of the DES survey area to the Large and Small Magellanic
Clouds (LMC and SMC, collectively MCs), these new objects appear good
candidates to have been members of the Greater Magellanic Galaxy,
imagined by \citet{lyndenbell76} as the tidally disrupted source of
material spanning half the sky, and subsequently associated with
various satellites of the Milky Way
\citep{lyndenbell76,lyndenbell82,lyndenbell95,donghia08}.

\begin{figure}
\includegraphics[width=\columnwidth]{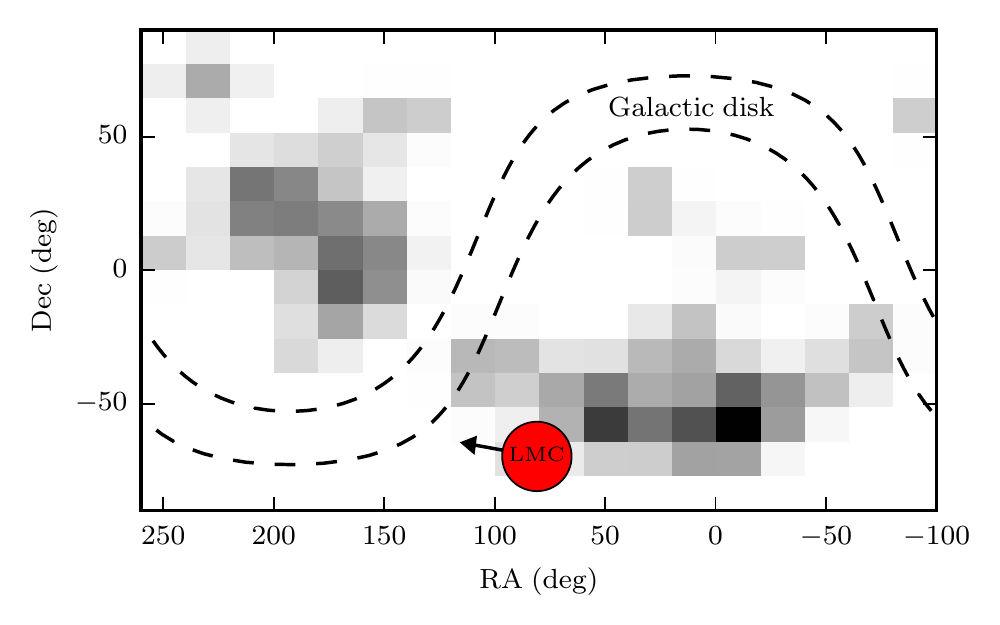}
\caption{Galactic satellite distribution in equatorial
  coordinates. The density distribution of all dwarf galaxies \citep[from
    the updated list of][]{mcconnachie12} within 350 kpc from the MW
  centre is shown as a greyscale 2D histogram. The sky is split into
  18 pixels along RA and 14 along Dec. Gaussian smoothing with a FWHM of
  1.7 pixels is applied. The red filled circle marks the current location
  of the LMC. The arrow represents the direction of the LMC proper
  motion corrected for the solar reflex. Lines (dashed) of constant
  Galactic latitude, $b=-10^{\circ}$ and $b=10^{\circ}$, are shown to
  highlight the location of the Galactic plane.}
\label{fig:dwarf_den}
\end{figure}

Figure~\ref{fig:dwarf_den} is a striking - even if somewhat
superficial - confirmation of the picture conjured by
\citet{lyndenbell76}.  Here, the on-sky density distribution of the
known Galactic dwarf satellites is shown to be dominated by a long and
broad band roughly aligned with the LMC's proper motion. Naturally, if
the association with the MCs was indeed real, the over-dense region
would be nothing but the leading (mostly above the Galactic plane) and
the trailing (mostly below) debris tails of the disrupting Magellanic
system. While this anisotropy has been alluded to on numerous
occasions recently \citep[e.g.][]{kroupa05,pawlowski12}, it is rather
difficult to assess its true significance as different parts of the
sky have so far been imaged to different depths. Therefore, to
mitigate the selection effects, we concentrate on the particular patch
of the sky where i) the satellite over-density is at its highest and
ii) deep and contiguous coverage is available, i.e. the DES footprint.

One of the main aims of this work is to establish the likelihood of association of the DES satellites with the MCs.
One approach in doing this has been to consider analogues of the LMC in cosmological zoom-in simulations of MW-like halos.
\citet{sales11} mapped the phase-space distribution of the subhaloes of one LMC analogue in the Aquarius simulations \citep{springel08} and, though unable to confirm any clear association with the \emph{then known} Galactic satellites, they did predict a concentration of satellites in the vicinity of the LMC if it has had only one MW pericentric passage.
\citet{deason15} generalised this by looking at 25 analogues found in the ELVIS simulations \citep{garrison14} finding a trend between LMC accretion time and the $z=0$ phase-space dispersion of its subhaloes.
Their comparison with 9 of the DES satellites suggested a recent ($<2$ Gyr) LMC accretion, and 2-4 satellites of likely Magellanic origin.

An alternative approach in establishing an association was taken by \citet{koposov15a}, who demonstrated that the region around the LMC is significantly over-dense with satellites under the assumption of an isotropic Galactic distribution.
This inconsistency of the DES satellites with an isotropic MW population is re-iterated in \citet{bechtol15b}, who further model the satellites' 3D spatial distribution as two spherical components centred on the LMC and SMC and an isotropic MW background.
This suggested that 20-30\% of all known satellite galaxies in the vicinity of the MW have been associated with the MCs.

The advantage of using zoom-in simulations is that they paint a self-consistent cosmological picture; the difficulty lies in finding an LMC analogue with the observed properties.
For example, both \citet{sales11} and \citet{deason15} necessarily omit the LMC's massive companion, the SMC.
Furthermore, a model constructed from an LMC analogue with the incorrect position and velocity cannot take full advantage of the satellites' phase-space information, e.g. \citet{deason15} give the probability of association as a function of 1D distance from the LMC rather than 3D position.
Conversely, the simple 3D model of \citet{bechtol15b} does not include any dynamical effects.

In this work we take a complementary approach, building a dynamical model which satisfies the observed kinematics.
We do this by performing fast orbit integrations in analytic potentials, in a similar way to \citet{nichols11}, extending the methods used in that work by using an improved dynamical friction model, including the MW's reflex motion due to the LMC - the importance of which was highlighted in \citet{gomez15} - and using the most recent measurements of the Clouds' proper motions \citep[][hereafter K13]{kallivayalil13}.
We also note the similarity of our method to \citet{yozin15}.

This paper is organised as follows. First, we give the details of our
simulation setup in Section~\ref{sec:sims}, and describe the initial conditions,
the methods of orbit integration and the effects of dynamical friction. We present the results of the spatial
distribution modelling of the Magellanic satellites in
Section~\ref{sec:3dmodel} and give the probability of belonging to the MCs and the
MW for each of the recently discovered objects. Section~\ref{sec:veloc} compares the observed kinematics with the
model predictions and forecasts the line-of-sight velocities for the
objects awaiting follow-up. Taking advantage of the full simulation
suite, we gauge the total mass of the LMC in
Section~\ref{sec:lmcmass}. Finally we discuss the main results and the
limitations of this work and conclude in Section~\ref{sec:conclusions}.

\section{Simulations}
\label{sec:sims}


We wish to approximate the $z=0$ distribution of satellites once associated with the MCs. To do this we assume that satellite dynamics are governed by the gravitational forces of the MW, LMC and SMC, i.e. we assume that the satellites can be treated as tracer particles in the combined potential of the three massive galaxies.
This is a reasonable assumption for the satellites we will be considering, which have luminosities at least 10 magnitudes lower than the SMC.
Having made this assumption we can model the $z=0$ Magellanic satellite distribution by:
\begin{enumerate}
\item initialising the MCs with their observed positions and velocities today
\item rewinding the orbits of the MW, LMC and SMC in their combined three-body potential, accounting for the effects of dynamical friction
\item introducing a distribution of tracer particles representing the satellites at some time in the past
\item integrating forward the orbits of the tracer particles in the three-body potential to today
\end{enumerate}

\subsection{Galaxy Models}
\label{ssec:galaxymodels}

\begin{table}
	\centering
	\caption{Galaxy model parameters. Mass units are $10^{10}\;M_\mathrm{\sun}$, length units are kpc.
	For the MW and SMC we take three values of $M_{200}$, for the LMC we take 15 values linearly spaced in the range shown.
	Non-numeric entries represent constraints which a parameter is chosen to satisfy: Plummer and Miyamoto-Nagai (MN) masses satisfy constraints from \citet{vdmarel14}$^{(\ast)}$, \citet{stanimirovic04}$^{(\star)}$ and \citet{mcmillan11}$^{(\dagger)}$.}
	\label{tab:galaxy_models}
	\begin{tabular}{lcccc}
	\hline\hline
	   &   & MW & LMC & SMC \\
	\hline
	\multirow{3}{*}{NFW} & $M_{200}$ & $\{50,100,200\}$ & [2,30] & $M_{200}^\mathrm{LMC}/\{3,5,10\} $ \\
	   & $c_{200}$ & $c(M)$ & $c(M)$ & $c(M)$ \\
	   & $r_\mathrm{trunc}$ & $R_{200}$ & $R_{200}$ & $R_{200}$ \\
	\hline
	\multirow{2}{*}{Plummer} & $M$ & \multirow{2}{*}{-} & $M_\mathrm{enc}^{(\ast)}$ & $M_\mathrm{enc}^{(\star)}$\\
	   & $a$ &   & 5 & 3 \\
	\hline
	\multirow{3}{*}{MN disk} & $M$ & $V_\mathrm{circ,\mathrm{\sun}}^{(\dagger)}$ & \multirow{3}{*}{-} & \multirow{3}{*}{-} \\
	   & $a$ & 3 &   &   \\
	   & $b$ & 0.28 &   &   \\
	\hline
	\end{tabular}
\end{table}

We represent the MW, LMC and SMC with two-component analytic potentials.
The models used are described here and their parameters are summarised in Table~\ref{tab:galaxy_models}.

The first component of each galaxy is a Navarro-Frenk-White (NFW) halo \citep{navarro97}.
We vary halo virial (i.e. $M_{200}$) masses, and assign a mass dependent concentration
\begin{equation}
\log_{10} c_{200} = 1.645 - 0.065 \log_{10} M_{200},
\label{eqn:cM}
\end{equation}
which is based on the $c(M)$ relation shown in \citet{sanchez14}.
Compared to the relation shown there, our Equation~(\ref{eqn:cM}) takes the mean concentration for a MW like halo but is less concentrated than a typical $M_{200} \sim 10^{9-10} M_\mathrm{\sun}$ halo.
This is required in order that our MC models do not have masses in tension with observations, as discussed below.

The second component of each galaxy is chosen to satisfy a mass constraint.
For the Milky Way we use a Miyamoto-Nagai (MN) disk \citep{miyamoto75} with scale lengths fixed at the values of \texttt{MWPotential2014} model described in \citet{bovy15}, and mass chosen to satisfy the \citet{mcmillan11} measurement of $V_\mathrm{circ}(R_\mathrm{\sun})=239\;\mathrm{km}\;\mathrm{s}^{-1}$ at the solar radius, $R_\mathrm{\sun}=8.29\;\mathrm{kpc}$.
For the LMC we use a Plummer model \citep{plummer11} of fixed size to satisfy the \citet{vdmarel14} constraint of $M(<8.7\;\mathrm{kpc})=1.7\times10^{10}\;M_\mathrm{\sun}$, and likewise for the SMC using the \citet{stanimirovic04} measurement of $M(<3.5\;\mathrm{kpc})=2.4\times10^{9}\;M_\mathrm{\sun}$.
We use spherical bulges rather than flattened disks here to avoid the complication of following disk orientations during galaxy interactions, justifying this departure from reality since we are interested in the dynamics of satellites in the outer haloes of these galaxies, far from their disks.

The only independent parameter varied in our models is the NFW $M_{200}$ mass of each galaxy, with ranges shown in Table~\ref{tab:galaxy_models}.
We adopt three values for the MW and SMC (hereafter referred to as Light, Medium and Heavy) and 15 values of LMC mass.
We focus our parameter grid this way since we expect the LMC to exert the dominant force at the position of many of the observed satellites.
With such a coarse grid in SMC mass, we parametrize it as some fraction of the LMC - rather than using absolute values - in order that it exerts similar influence on the satellite dynamics over the whole LMC mass range.

For the LMC's virial mass, our lower bound of $2\times10^{10}\;M_\mathrm{\sun}$ is only slightly larger than the known dynamical mass \citep{vdmarel14} while our upper bound of $3\times10^{11}\;M_\mathrm{\sun}$ is motivated by the \citet{penarrubia15} measurement of the mass of the LMC.
This measurement is based on the LMC's influence on the barycentre of the Local Group.
We take 15 values of the LMC mass spaced linearly in this range.
The factor of four separating our lightest and heaviest MWs reflects our uncertainty in this quantity, bounded by values suggested by \citet{gibbons14} and \citet{boylankolchin13}.
With little prior knowledge of the LMC:SMC mass ratio we take our smallest ratio to be the 3:1 ratio of their \textit{V}-band luminosities \citep{deVaucouleurs91}, and arbitrarily take 10:1 as our most extreme case.

Motivated by the identification of the \emph{splash-back radius} of cold dark matter haloes beyond which a sharp density drop is observed \citep{more15}, we truncate the density profiles of our galaxy models. This physically motivated truncation also gives more realistic orbits at the large distances we will consider and gives our NFW profiles a finite mass.
We will require a truncation which gives a continuous density slope (as discussed in Section~\ref{ssec:tracers}) hence adopt an exponential truncation following \citet{springel99}, 
\begin{equation}
\label{eqn:trunc}
\rho_T(r)=
\begin{cases}
	\rho(r) & \text{if } r \leq r_T\\
	\rho(r_T) \left(\frac{r}{r_T}\right)^\epsilon \exp\left( - \frac{r-r_T}{r_D} \right) & \text{if } r > r_T
\end{cases}
\end{equation}
where $\rho(r)$ is the un-truncated density profile, $r_T$ is the truncation radius, $r_D$ is a decay radius and $\epsilon$ is chosen to give a continuous slope.
The exact value of the splash-back radius is found to lie in the range $0.8-1.5 R_{200}$ and is a function of the halo's accretion history \citep{more15}.
For simplicity, following \citet{kazantzidis04}, we take a fixed value of $r_T = R_{200}$ and decay radius $r_D = 0.1 R_{200}$.

\subsection{Kinematics}
\label{ssec:kinematics}

\begin{table}
	\centering
	\caption{Kinematics of the MCs. LMC values are adopted from \citet{vdmarel14}, SMC from \citetalias{kallivayalil13}.}
	\label{tab:kinematics}
	\begin{tabular}{lcc}
	\hline\hline
	   & LMC & SMC \\
	  \hline
	  RA (deg) &  $79.88\pm0.83$ & $16.25\pm0.20$ \\
	  Dec. (deg) &  $-69.59\pm0.25$ & $-72.42\pm0.20$ \\
	  DM  &  $18.5\pm0.1$ &  $18.99\pm0.1$ \\
	  $v_{LOS}$ (km $\mathrm{s}^{-1}$) & $261.1\pm2.2$ & $145.6\pm0.60$ \\
	  $\mu_W$ (mas $\mathrm{yr}^{-1}$) & $-1.895\pm0.024$  & $-0.772\pm0.063$ \\
	  $\mu_N$ (mas $\mathrm{yr}^{-1}$) & $0.287\pm0.054$ & $-1.117\pm0.061$ \\
	  \hline
	\end{tabular}
\end{table}

The kinematic observations we use for the MCs are shown in Table~\ref{tab:kinematics}.
Values for the SMC are taken from \citetalias{kallivayalil13}.
For the LMC we use \citet{vdmarel14} over \citetalias{kallivayalil13} since the former solves for the proper motion of the LMC centre-of-mass by simultaneously fitting both proper motion and radial velocity measurements (PM + Old $v_\mathrm{LOS}$ Sample from their Table 1).
This is more consistent than the approach taken in \citetalias{kallivayalil13}, however the difference in the final solution is small.

Assuming Gaussian errors on the all values shown in Table~\ref{tab:kinematics}, we draw 100 samples from the joint LMC/SMC distribution and convert them to Galactocentric Cartesian co-ordinates which form the initial conditions for the MCs in our simulations.
While the MW is initialised at rest at the origin, it is free to respond to the gravity of the MCs at later times.

Throughout this work, for all co-ordinate transformations we assume the \citet{mcmillan11} measurement of $V_\mathrm{circ}(R_\mathrm{\sun})=239\;\mathrm{km}\;\mathrm{s}^{-1}$ at the solar radius, $R_\mathrm{\sun}=8.29\;\mathrm{kpc}$, and the \citet{schonrich10} measurement of the sun's peculiar motion, $(U,V,W)_\mathrm{\sun}=(11.1,12.24,2.25)$ km s$^{-1}$.

\subsection{Dynamical Friction}
\label{ssec:dynfric}

We include the effect of dynamical friction on the trajectories of our three massive galaxies using Chandrasekhar's formula \citep{binney08},
\begin{equation}
\label{eqn:dynfric}
\frac{\mathrm{d}\mathbf{v}}{\mathrm{d}t} = - \frac{4 \pi G^2 M \rho \ln \Lambda}{v^2} \left[ \mathrm{erf}(X) - \frac{2X}{\sqrt{\pi}} e^{-X^2}\right] \frac{\mathbf{v}}{v},
\end{equation}
where $M$ is the mass of a satellite moving with velocity $\mathbf{v}$ in the density field $\rho$ of some host body, $X = v / (\sqrt{2}\sigma)$ where $\sigma$ is the local 1D velocity dispersion of the host, and $\ln \Lambda$ is the Coulomb logarithm.

We make the following choices for these quantities. For $\rho$ we use the total density of all components of the host, while for $\sigma$ we use the velocity dispersion of its NFW halo, using the approximate form given by equation 6 of \citet{zentner03}.
For satellite mass we use the $M_{200}$ of its dark halo.
For the Coulomb logarithm we use
\begin{equation}
\ln \Lambda = \ln \left( \frac{r}{\epsilon} \right),
\end{equation}
where the use of instantaneous separation, $r$, in place of the maximum impact parameter was shown by \citet{hashimoto03} to better reproduce the orbital decay time-scale compared to using some fixed value.

The length $\epsilon$, interpreted as the minimum impact parameter effectively taking part in dynamical friction, depends on the satellite's density profile \citep{white76}.
The scaling $\epsilon=1.6 r_s$, where $r_s$ is the satellite's scale length, has previously been used when modelling the LMC as a Plummer sphere \citep[e.g.][K13]{besla07,sohn13}.
This is derived analytically in \citet{hashimoto03}, however they go on to show that $\epsilon=1.4 r_s$ better fits their N-body simulation.
Given this, and the departure of our LMC models from Plummer profiles, we choose to calibrate $\epsilon$ against N-body simulations of MW/LMC interactions.

\begin{figure}
\includegraphics[width=\columnwidth]{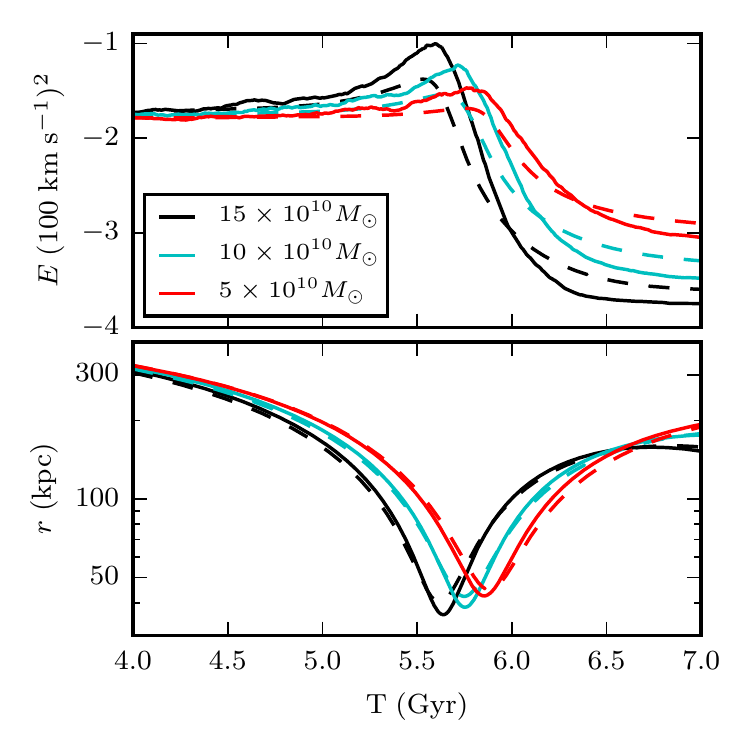}
\caption{
Calibrating Chandrasekhar dynamical friction.
We show the evolution around pericenter of orbital energy (\textit{top}) and separation (\textit{bottom}) of three LMCs with NFW virial masses as shown in the legend.
\textit{Solid lines} are the results of N-body simulations.
\textit{Dashed lines} show the best fitting orbits found by varying $\epsilon$ in the Coulomb logarithm.
}
\label{fig:dynfric}
\end{figure}

To do this, we ran three N-body simulations using the N-body part of \textsc{gadget-3} which is similar to \textsc{gadget-2} \citep{springel05}. The LMCs were each made up of an NFW halo and a Plummer bulge. The NFW components had a masses of 5, 10 and 15 $\times 10^{10} M_\mathrm{\sun}$ with concentrations of 9.2, 8.66, 9.35 respectively. The Plummer bulges all had $r_s=6.67$ kpc with masses of $1.98, 1.41, 1.01 \times 10^{10} M_\odot$ (from lightest to heaviest LMC) necessary to satisfy the LMC mass constraint in \citet{vdmarel14}. These were evolved in a halo similar to our Medium MW, with $M_{200} = 10^{12} M_\odot$ and $c_{200} = 12$. We did not include the MW disk since it only makes a minor contribution to the density, and hence to dynamical friction, in the regions of interest. 
The NFW halo of the LMC was modelled with $10^5$ particles with a softening of 200 pc, the Plummer bulge of the LMC was modelled with $10^4$ particles with a softening of 890 pc \citep[the optimal Plummer softening from][]{dehnen2001}. The NFW halo of the MW was modelled with $10^6$ particles and a softening of 200 pc. The LMC and MW were initialised using the procedure described in \cite{kazantzidis04}. For the NFW profiles, we used a truncation radius of $R_{200}$ and a decay radius of $r_D = 0.1 R_{200}$, extending the profile to a hard cutoff at $R_{200} + 3r_D$. For the Plummer profiles embedded in the NFWs we used a hard truncation radius of $R_{200} + 3 r_D$. We note that we did not adiabatically contract the NFW when generating the initial conditions so that the simulations could be directly compared against the model described in Section~\ref{ssec:orbint}.

The LMCs were initialised at $500$ kpc with an entirely tangential velocity of $32.7$ $\mathrm{km}\;\mathrm{s}^{-1}$.
A tracer particle on this orbit would have pericenter at 50 kpc, similar to the real LMC, after 5.6 Gyr. The MW was initialised at the origin with zero velocity. Each simulation was run for 7 Gyr and the orbits of the LMC and MW were computed by locating their density peaks at each time. 

We fit the N-body orbits with MW/LMC orbits calculated using analytic potentials as described in Section~\ref{ssec:orbint}, including a dynamical friction force on the LMC and allowing $\epsilon$ to vary.
Note that in the analytic potential orbits we include the reflex motion of the MW, the exclusion of which artificially enhances dynamical friction \citep{white83}.
We perform the orbit fits in $(E,t)$ space, defining the orbital energy of the LMC, 
\begin{equation}
E = \frac{1}{2}|\mathbf{v}_\mathrm{LMC}-\mathbf{v}_\mathrm{MW}|^2 + \Phi_\mathrm{MW}(|\mathbf{r}_\mathrm{LMC}-\mathbf{r}_\mathrm{MW}|),
\label{eqn:orbitalenergy}
\end{equation}
and finding the $\epsilon$ which minimises the chi-squared error between the N-body and analytic potential ($\Phi$) orbit,
\begin{equation}
\chi^2 = \sum\limits_{t} \frac{[E_\mathrm{Nbody}(t) - E_{\Phi}(t,\epsilon)]^2}{|E_{\Phi}(t,\epsilon)|}.
\end{equation}

Figure~\ref{fig:dynfric} shows the N-body and best-fitting analytic potential orbits for the three LMC models.
We show both energy and separation as a function of time either side of pericenter.
Note the energy of Equation~(\ref{eqn:orbitalenergy}) is defined in the non-inertial MW frame, resulting in its non-conservation, i.e. the slow rise seen between 4-5.5 Gyr.
The sharp drop in energy between 5.5-6.5 Gyr is the effect of dynamical friction.
For each of the three LMCs, an $\epsilon$ is found which satisfactorily reproduces the loss of orbital energy and also well describes the evolution in orbital separation.
We summarise the dependence of the optimal $\epsilon$ on the NFW scale length $r_s$ as
\begin{equation}
\epsilon=
\begin{cases}
	2.2 r_s - 14 & \text{if } r_s \geq 8 \; \mathrm{kpc} \\
	0.45 r_s     & \text{if } r_s < 8  \; \mathrm{kpc}
\end{cases}
\end{equation}
where scale lengths outside the range $8<r_s / \mathrm{kpc}<14$ are extrapolations beyond our simulated range.

We note that the MW and LMCs used in our N-body calibration of dynamical friction do not exactly match the ones we will use in our model of the LMC satellites, i.e.  those in Table \ref{tab:galaxy_models}. We have re-run the N-body simulations with the MW and LMCs described in Table \ref{tab:galaxy_models} and find a small change the dependence of $\epsilon$ on $r_s$. To check the sensitivity of our results to our dynamical friction prescription, we re-ran part of the analysis in Section \ref{sec:3dmodel} using the updated $\epsilon$ and found no significant change in the LMC and SMC satellite properties. 

While Figure \ref{fig:dynfric} shows that our dynamical friction prescription can reasonably reproduce our N-body results, the application of Chandrasekhar's formula for soft and massive galaxies is known to be fraught with uncertainties \citep{white83}.
We acknowledge a number of concerns regarding our implementation: our extrapolation for $\epsilon$ may not be valid, while our N-body simulations only test one MW host and one LMC orbit.
It is beyond the scope of this work to fully address all of these issues, however at least for orbits with a single pericentric crossing, our implementation of Chandrasekhar's formula should capture the magnitude of the loss of orbital energy.

\subsection{Satellite tracers}
\label{ssec:tracers}

\begin{figure}
\includegraphics[width=\columnwidth]{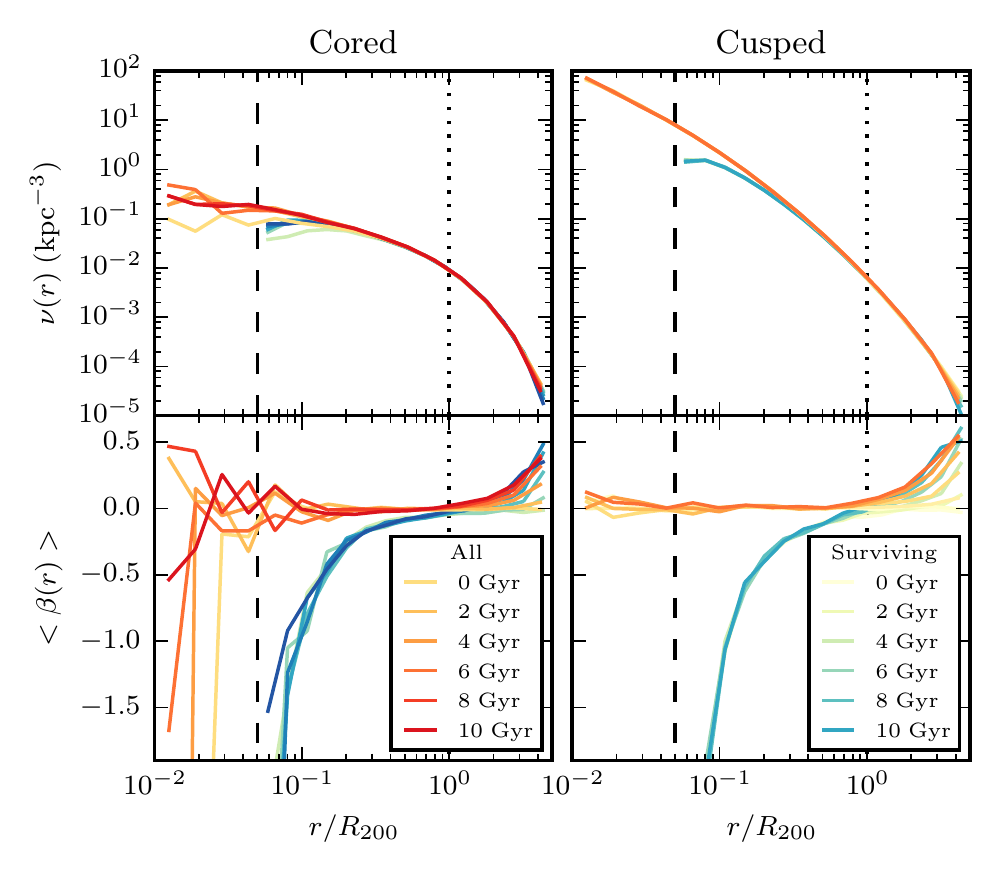}
\caption{Evolution in density profile/anisotropy parameter (top/bottom
  panels) for our cored/cusped (left/right panels) when evolved in
  their isolated host galaxy.  Red shades show the evolution of the
  entire population, blue shades show only those satellites which
  survive till present day.  Dashed lines show the destruction radius,
  dotted lines show $R_{200}$. Note two departures from orbital
  isotropy. First, from the onset, due to the density tapering at
  large distances, a radial bias exists at $r>2R_{200}$. Second, when
  satellites are removed within the designated destruction radius,
  a tangential bias sets in at small distances from the MCs' centres.}
\label{fig:checkICs}
\end{figure}

Our aim is to model the $z=0$ distribution of satellites once associated with either the LMC or SMC.
We define ``satellite once associated with'' to mean a satellite galaxy which at some time in the past was within the virial radius of its host and has survived as a self-bound entity to $z=0$.
By restricting to satellite galaxies we can turn to the properties of subhaloes in cosmological simulations to guide our assumptions.

To determine when to introduce our satellites, we ask at what time we expect the MCs to have assembled the bulk of their subhalo populations.
\citet{wetzel09} showed that subhalo occupation number peaks at $z\sim2.5$, albeit for mass scales an order of magnitude larger than those we consider for the MCs, with a trend for less massive haloes to have peak subhalo occupation at earlier times.
We therefore take the earlier time of 11.5 Gyr ago, roughly corresponding to $z\sim3$, to introduce tracer particles.

For the satellites' initial spatial distribution we use the \citet{ludlow09} determination of the $z=0$ radial number density profile of all subhaloes of a MW-like host.
This is given by an Einasto profile,
\begin{equation}
\label{eqn:einsato}
\nu(r) \propto \exp \left\{- \frac{2}{\alpha} \left(\left(\frac{r}{r_{-2}}\right)^\alpha-1\right) \right\},
\end{equation}
with parameters $r_{-2}=R_{200}$ and $\alpha=0.8$.
We adopt these values as our ``\textit{cored}'' satellite distribution
There are a number of uncertainties however, in our use of this profile, which was calculated for MW like host haloes at $z=0$, whereas we initially wish to model lower mass hosts at $z\sim3$.
We therefore also adopt a second, ``\textit{cusped}'' satellite distribution with Einasto parameters $r_{-2}=0.1 R_{200}$ and $\alpha=0.2$, which roughly follows the distribution of dark matter in the NFW halo, to allow us to gauge the effect of this systematic uncertainty on our results.
The cusped profile is also loosely motivated by the suggestion from hydrodynamic simulations that dwarf galaxies exist in a biased population of subhaloes preferentially concentrated around their host \citep{sawala14}.
The Einasto $\alpha$ parameters of our cored and cusped profiles bracket that found in the high resolution collisionless simulations of \citet{springel08}.

For each MC model described in Section~\ref{ssec:galaxymodels}, we initially sample $10^5$ radial positions from both the cored and cusped distributions, out to a maximum radius of $5 R_{200}$.
Allowing for initial radii greater than $R_{200}$ is a way to include subhaloes which have not yet been accreted at $z=3$ but are on orbits likely to enter within $R_{200}$ at some time before today.
Of course, evolving in the external potential of the MW, many of these budding Magellanic satellites will never enter the virial radius of their prospective host.
We explicitly exclude these spurious satellites from our results by insisting that a tracer particle must at some point during its orbit enter within $R_{200}$ of its host if it is to be counted as a satellite.

For both the cored and cusped initial density profiles, and for each MC model, we generate the corresponding ergodic distribution function by performing an Eddington inversion \citep{eddington16,binney08} and checking that the distribution function is nowhere negative.
We note here that the decision to use an exponential truncation, given by Equation~(\ref{eqn:trunc}), over a hard truncation was driven by the requirement of a continuous density slope when performing the Eddington inversion.
We sample velocities from the distribution function using the acceptance-rejection technique \citep{kuijken94}, then randomise angles to give initial conditions for a spherical, isotropic, non-rotating tracer population of satellites.

Figure~\ref{fig:checkICs} shows an example of the two initial distributions, calculated for our $10^{11} M_\mathrm{\sun}$ LMC model.
We test the stability of our initial conditions by integrating orbits for 10 Gyr, and plotting the evolution in density and anisotropy profiles of the tracer population at 2 Gyr intervals shown by the sequence of coloured lines.
Both cored and cusped density profiles are sufficiently stable, with only some minor evolution observed at small radii for the cored profile.
Within $R_{200}$ both populations remain isotropic, however at radii $r>2R_{200}$ a radial bias quickly develops.
This is due to the lack of tracers beyond the imposed radial cut off of $5R_{200}$ needed to replenish high angular momentum orbits at these distances.

To satisfy the criterion of survival until $z=0$, we require that a satellite must not enter within some threshold ``destruction radius'' of either MC during its lifetime to avoid tidal disruption.
We pick radii of 5 kpc for the LMC and 3 kpc for the SMC, which are roughly double their respective half-light radii, recently re-measured in \citet{torrealba16}.
We assume that this is a reasonable lower bound for the distance at which tidal disruption occurs, though in reality this will depend on the satellite's mass distribution, with tidal disruption occurring at larger distances for more extended satellites.

The sequence of blue lines in Fig.~\ref{fig:checkICs} show the density and anisotropy evolution when we only retain satellites which have survived until the present day.
Unsurprisingly, within $R_{200}$ these satellites have a tangential bias, since satellites on very radial orbits are more likely to be tidally disrupted.
This effect is also observed in cosmological simulations \citep{ludlow09}.

\subsection{Orbit integrator}
\label{ssec:orbint}

Now that we have described our galaxy models and their satellite distributions, we give the details of our three-body orbit integration scheme.

Each massive galaxy $g \in \{\mathrm{M,L,S}\}$ (where $M,L,S$ stand for MW, LMC, SMC) is represented by a potential model $\Phi_g(\mathbf{x})$, position $\mathbf{x}_g(t)$ and velocity $\mathbf{v}_g(t)$.
We approximate the acceleration $\mathbf{a}_{gh}(t)$ of galaxy $g$ due to the force of galaxy $h$ as the acceleration of a point mass at the position of galaxy $g$ in the potential of host galaxy $h$:
\begin{equation}
\mathbf{a}_{gh}(t) = - \nabla \Phi_h(\mathbf{x}_g(t) - \mathbf{x}_h(t)).
\end{equation}
Note that this approximation is good when the galaxies are well separated but breaks down when significant fractions of their mass distributions are coincident.
Letting $\mathbf{a}_\mathrm{gh}^\mathrm{DF}$ be the dynamical friction drag on galaxy $g$ due to its motion in galaxy $h$ of Equation~(\ref{eqn:dynfric}), the total accelerations of the three bodies are given by
\begin{align}
&\mathbf{a}_\mathrm{M}(t) = \mathbf{a}_\mathrm{ML}(t) + \mathbf{a}_\mathrm{MS}(t),\\
&\mathbf{a}_\mathrm{L}(t) = \mathbf{a}_\mathrm{LM}(t) + \mathbf{a}_\mathrm{LS}(t) + \mathbf{a}_\mathrm{LM}^\mathrm{DF}(t),\\
&\mathbf{a}_\mathrm{S}(t) = \mathbf{a}_\mathrm{SL}(t) + \mathbf{a}_\mathrm{SL}(t) + \mathbf{a}_\mathrm{SM}^\mathrm{DF}(t) + \mathbf{a}_\mathrm{SL}^\mathrm{DF}(t),
\end{align}
where we have chosen to include both the dynamical friction of the MW on both MCs, and the dynamical friction of the LMC on the SMC.
We integrate the equations of motion using a symplectic leapfrog scheme \citep{springel01} and adaptive time-steps
\begin{equation}
\Delta t = 0.01 \; \min_{g,h} \left( \sqrt{\frac{|\mathbf{x}_g - \mathbf{x}_h|}{|\mathbf{a}_{gh}|}} \right),
\end{equation}
i.e. a hundredth of the shortest dynamical time-scale between pairs of galaxies.
To calculate forces we use the minimal \texttt{C} implementations of potentials included in the \texttt{Galpy} software package \citep{bovy15}.

We rewind the three-body orbits to $t_0 = 11.5$ Gyr.
At this time we introduce a tracer particle $T$ with initial conditions
\begin{align}
&\mathbf{x}_\mathrm{T}(t_0) = \mathbf{x}_\mathrm{init}^\mathrm{(L/S)} + \mathbf{x}_\mathrm{(L/S)}(t_0),\\
&\mathbf{v}_\mathrm{T}(t_0) = \mathbf{v}_\mathrm{init}^\mathrm{(L/S)} + \mathbf{v}_\mathrm{(L/S)}(t_0),
\end{align}
where $\mathbf{x}_\mathrm{init}^\mathrm{(L/S)}$ and $\mathbf{v}_\mathrm{init}^\mathrm{(L/S)}$ are initial conditions sampled from one of our populations of tracer satellites, which we centre on the LMC or SMC host as appropriate.
Tracer $T$ is integrated forward in the non-static three-body potential, with acceleration given by
\begin{equation}
\mathbf{a}_\mathrm{T}(t) = \mathbf{a}_\mathrm{TM}(t) + \mathbf{a}_\mathrm{TL}(t) + \mathbf{a}_\mathrm{TS}(t)
\end{equation}
where we update the positions of the MW, LMC and SMC by linearly interpolating the stored results of the backwards integration.

We integrate the tracer orbits forward to today subject to the following rules.
If the tracer ever ventures within the destruction radius of either the LMC or SMC, we say it has been tidally destroyed. If the tracer never enters within $R_{200}$ of its host, we do not count it as a true satellite. If a true satellite survives until today we output its position and velocity, continuing this process until we have output 5000 tracers for each set of our model parameters.

To summarise, of our grid of models comprises
\begin{itemize}
\item 135 galaxy mass model combinations \\ \indent i.e. 3 MWs $\times$ 15 LMCs $\times$ 3 SMCs
\item 100 sampled kinematics for the MCs
\item 4 satellite populations \\ \indent i.e. (cored, cusped) $\times$ (LMC, SMC)
\item 5000 tracers output for each satellite population
\end{itemize}
The total output is $2.7 \times 10^8$ tracer particles.
Accounting for tracer particles which enter the destruction radius or never enter within $R_{200}$, this requires the calculation of $3.8 \times 10^8$ tracer orbits, with a typical run-time of $5\times10^{-4}$ s per orbit.

\subsection{Orbit Comparison}

Before looking at the satellite distributions, we compare our MW+LMC+SMC orbit integrations with those of \citetalias{kallivayalil13}.
Our method differs from the one employed in that work in numerous ways.
In order of importance, these include our use of:
\begin{enumerate}
\item a non-static MW, free to respond to the LMC's gravity
\item LMC dynamical friction on the SMC
\item different potential models
\item different $\epsilon$ in Chandrasekhar dynamical friction
\item \citet{vdmarel14} LMC kinematics
\end{enumerate}
For a subsample of our galaxy mass models, we sample $10^3$ values of the MC kinematics and perform backward orbit integrations for 10 Gyr, storing various quantities as discussed below.

\begin{figure}
\includegraphics[]{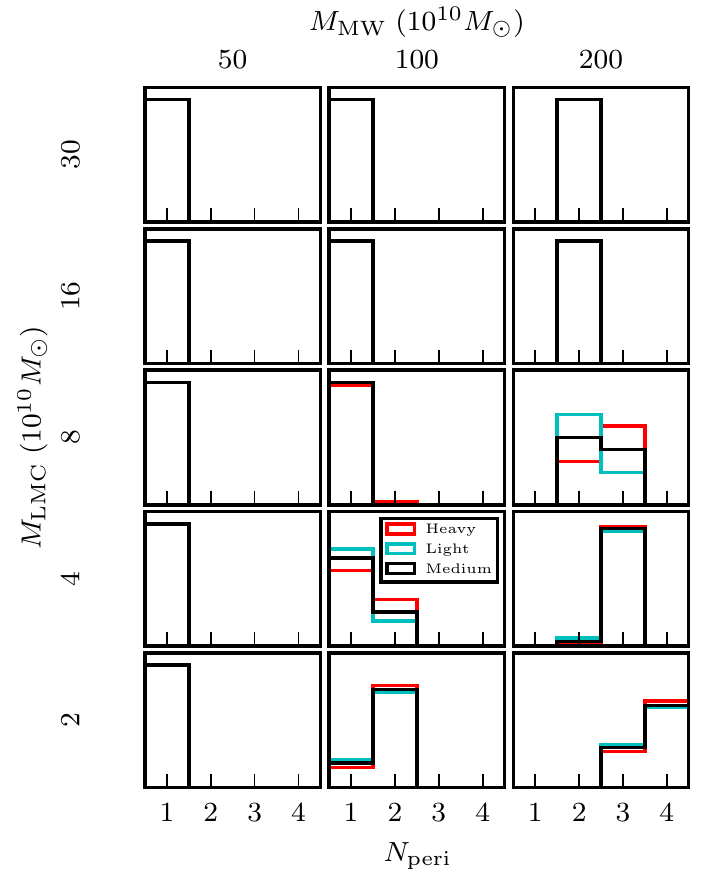}
\caption{
The number of MW-LMC pericenters in the past 10 Gyr.
Each panel shows the distribution of $N_\mathrm{peri}$ for a particular MW/LMC mass combination (masses increasing rightwards/upwards respectively) and, where visible, broken down by SMC mass as colours shown in the legend.
}
\label{fig:nperi}
\end{figure}

Figure~\ref{fig:nperi} shows the distribution of the number of
pericenters undergone between the MW and LMC in the last 10 Gyr, to be
compared with Fig. 9 of \citetalias{kallivayalil13}.  The results are
qualitatively similar: $N_\mathrm{peri}$ increases as we increase the
mass of the MW and decrease the mass of the LMC.  Looking at the exact
numbers predicted, however, we typically predict one more pericenter
than \citetalias{kallivayalil13} for all LMCs in our $2\times10^{12}
M_\mathrm{\sun}$ MW, and for the lightest LMC in our $1\times10^{12}
M_\mathrm{\sun}$ MW.  This is due to our use of a free MW which, as
discussed in \citet{gomez15}, results in shorter LMC orbital
periods. Importantly, we have checked that artificially pinning the MW
to the origin reproduces the results of \citetalias{kallivayalil13}.

\begin{figure}
\includegraphics[height=3.5in]{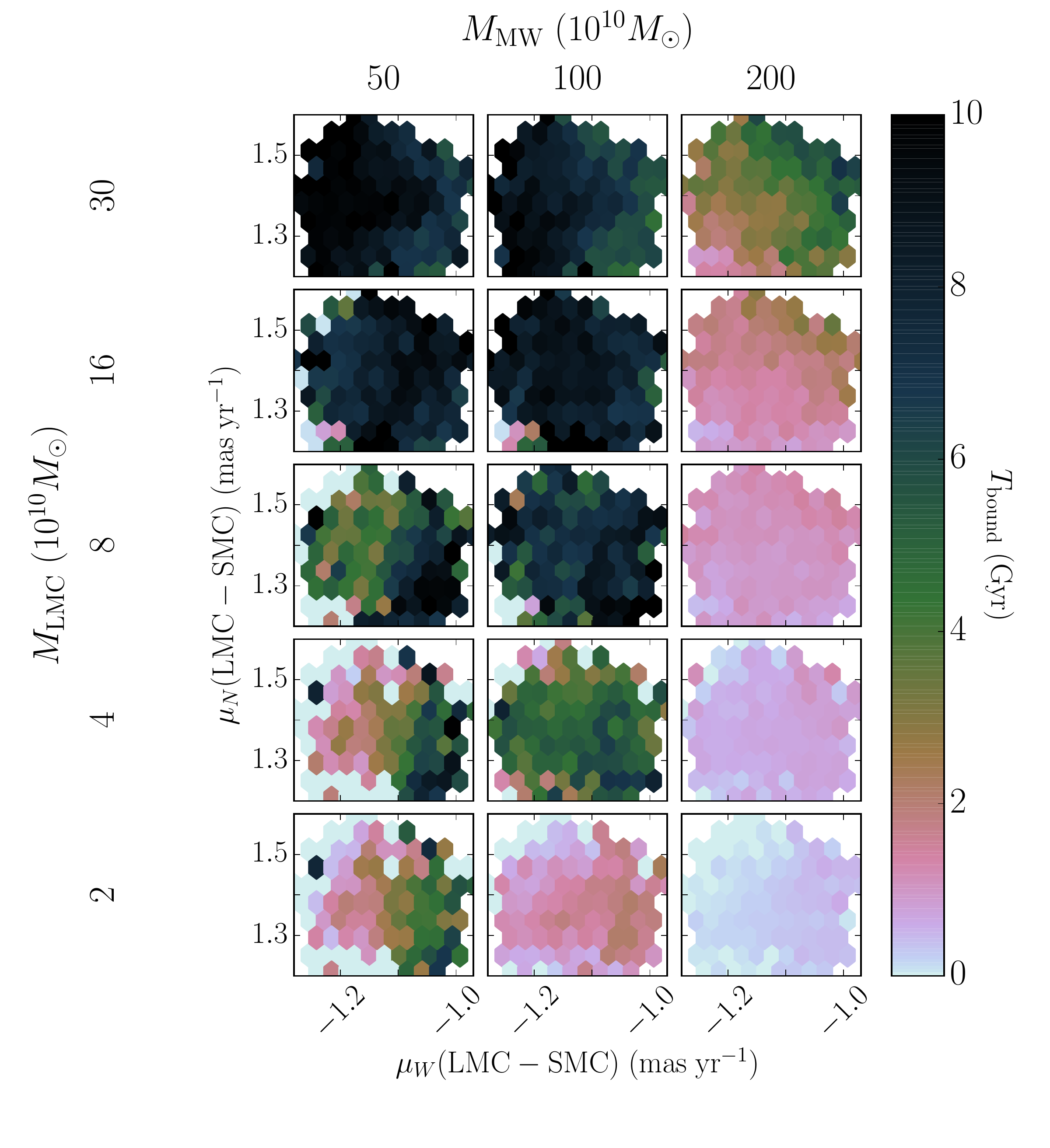}
\caption{
Time the LMC and SMC remained bound during their orbit.
Each panel shows results as a function of the relative LMC-SMC proper motion, for a particular MW/LMC mass combination, where we have only considered our light SMC model.
Colours represent the bound time averaged for all orbits in that proper motion cell.
}
\label{fig:t_bnd}
\end{figure}

Figure~\ref{fig:t_bnd} shows the amount of time the LMC and SMC were bound during their orbit, defined as in \citetalias{kallivayalil13} as the longest duration that the SMC's orbital energy relative to LMC is negative.
Results are shown for our light SMC model.
The panel with mass combinations most directly comparable to Fig. 14 of \citetalias{kallivayalil13} is shown in the lower right.
In both our and \citetalias{kallivayalil13} distributions, the longest bound states are found near $(\mu_W,\mu_N)(\mathrm{LMC}-\mathrm{SMC}) \approx (-1.0,1.5)$.
Unlike \citetalias{kallivayalil13} however, for this particular combination of masses, we do not find any solutions bound for $>2$ Gyr, which we attribute to our inclusion of LMC-SMC dynamical friction.
We do find that longer binary states are easily achievable for lower MW or higher LMC masses, in broad agreement with Fig. 12 of \citetalias{kallivayalil13}.

\section{3D Satellite Distribution Model}
\label{sec:3dmodel}

Having described our simulations, we will now explain how we use them to model the 3D spatial distribution of the satellites observed in DES and then compare our model against the data.

\subsection{Data \& Selection Function}
\label{ssec:data}

Table~\ref{tab:sats} lists the positions of 14 of the newly discovered satellites in the DES survey \citep{koposov15a,bechtol15b,kim15b} which we compare to our model.
The red points in the top left panel of Fig.~\ref{fig:mwmod} shows their luminosity as function of distance, the right panel their on-sky positions.

\begin{table}
	\centering
	\caption{Properties of satellites in the DES footprint.
	Values are taken from (1) \citet{koposov15a} (2) \citet{bechtol15b} (3) \citet{kim15b}.
	}
	\label{tab:sats}
	\begin{tabular}{lccccc}
	\hline\hline
	Name & $l$  & $b$  & $D_\mathrm{\sun}$ & $M_V$  & Ref.\\

		 & (deg) & (deg) & (kpc) & (mag) & \\
	\hline
	Columba 1 & 231.6 & -28.9 & 182 & 	-4.5		& (2) \\
	Grus 1 & 338.7 & -58.2 & 120 & 		-3.4		& (1) \\
	Grus 2 & 351.1 & -51.9 & 53 & 		-3.9		& (2) \\
	Horologium 1 & 271.4 & -54.7 & 79 & -3.4		& (1) \\
	Horologium 2 & 262.5 & -54.1 & 78 & -2.6 		& (3) \\
	Indus 2 & 354.0 & -37.4 & 214 & 	-4.3		& (1) \\
	Phoenix 2 & 323.7 & -59.7 & 83 & 	-2.8		& (1)\\	
	Pictoris 1 & 257.3 & -40.6 & 115 &  -3.1		& (1) \\
	Reticulum 2 & 266.3 & -49.7 & 30 & 	-2.7		& (1) \\
	Reticulum 3 & 273.9 & -45.6 & 92 & 	-3.3		& (2) \\
	Tucana 2 & 328.1 & -52.3 & 58 & 	-4.3		& (1) \\
	Tucana 3 & 315.4 & -56.2 & 25 & 	-2.4		& (2)\\
	Tucana 4 & 313.3 & -55.3 & 48 & 	-3.5		& (2) \\
	Tucana 5 & 316.3 & -51.9 & 55 & 	-1.6		& (2) \\
	\hline
	\end{tabular}
\end{table}

To compare our model with the observed satellites requires knowledge of the DES selection function.
The estimated detection efficiencies of the satellites vary between 1.0 and 0.2 \citep{bechtol15b} dependent on satellite distance, luminosity and size.
Given the systematic uncertainties in determining these parameters (e.g. \citet{koposov15a} and \citet{bechtol15a} give the luminosity of Reticulum 2 as $M_V=-2.7\pm0.1$ and $-3.6\pm0.1$ respectively), there will be large uncertainties in reported detection efficiencies.
Guided by the sharp boundary between detection and non-detection in efficiency maps calculated for the Sloan Digital Sky Survey (SDSS) \citep{koposov08}, we circumvent this uncertainty by assuming that detection is a binary decision.
Furthermore, we assume that this decision is determined entirely by distance and luminosity, omitting any dependence on size.
In effect, this means that our results will not account for any hypothetical population of undetectably-low surface-brightness dwarfs.
Our criteria for detectability DES is thus
\begin{align}
M_V &< M_{V,\mathrm{lim}}^\mathrm{DES}(D_\mathrm{\sun}) \nonumber \\
	&= (1.45 - \log_{10} (D_\mathrm{\sun}/ \mathrm{kpc})) / 0.228.
	\label{eqn:DES_selection}
\end{align}
The slope of this relation is assumed equal to that calculated for SDSS \citep[][i.e. Equation (\ref{eqn:dr5})]{koposov09} whilst the intersect is chosen such that all DES satellites lie above this threshold, as shown in the top left panel of Fig.~\ref{fig:mwmod}.

In compiling Table~\ref{tab:sats}, we have excluded all likely or known globular clusters, along with the distant dwarf galaxy candidate Eridanus 2 which lies beyond the distance range of our Magellanic satellite distributions ($D_\mathrm{\sun}=380$ kpc).
We also exclude the faintest dwarf candidate Cetus 2, at whose luminosity ($M_V=0$ mag) our census of satellites is woefully incomplete.
The inclusion of Cetus 2 would therefore affect our results in a manner highly dependent on extrapolations of the luminosity function well beyond current understanding.

\subsection{Method}

We now describe the steps we take to compare our simulated MC satellites to the observed distribution.

\subsubsection{MW Background}
\label{ssec:mwmod}

\begin{figure*}
\includegraphics[width=\textwidth]{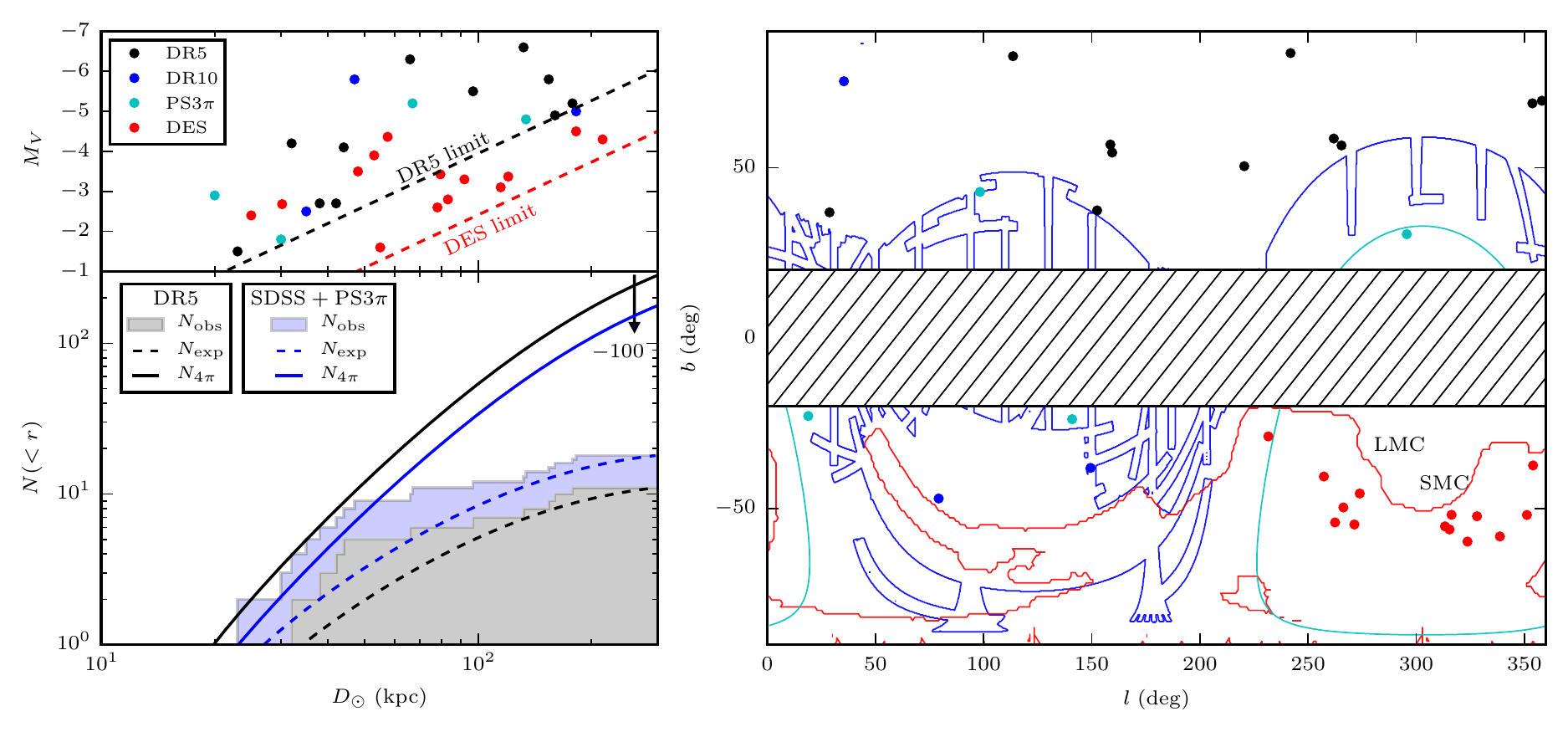}
\caption{
Properties of known $-7<M_V<-1$ dwarfs.
\textbf{Top left}: as a function of distance we show (i) satellite luminosity (coloured dots), (ii) the detection limit of DR5 (dashed black line), (iii) our assumed detection limit of DES (dashed red line).
\textbf{Right}: in Galactic co-ordinates (i) satellite positions (coloured dots), (ii) survey footprints (solid lines), (iii) the positions of the MCs (as text), (iv) $20^\circ$ masked either side of the Galactic disc.
The right and top left panels are colour coded according to the legend in the top left panel, representing the survey in which a satellite was discovered.
\textbf{Bottom left}: cumulative number counts of dwarfs as a function of distance. We show (i) the observed distribution from DR5 (filled grey histogram) (ii) our modelled MW distribution $N_\mathrm{exp}$ with normalisation $N_\mathrm{DR5}$ as it would be observed given the DR5 detection threshold (dashed black line), (iii) the predicted number of all dwarfs over the whole sky $N_{4\pi}$ assuming a normalisation $N_\mathrm{DR5}$ (solid black line).
The same quantities for the complete SDSS and PS$3\pi$ surveys are shown in blue.
All lines assume a luminosity function with $\gamma=1.9$ and the virial radius $R_{200}=210$ kpc of our Medium MW model.
}
\label{fig:mwmod}
\end{figure*}

To allow for the possibility that some of the observed satellites do not have a Magellanic origin, we include an isotropic background component of MW satellites in our model.
The radial profile of this component is fixed to the $\Lambda$CDM prediction of the distribution of subhaloes, while we allow its normalisation to vary in a range informed by the number of dwarfs observed in surveys prior to DES.
Here we describe these choices.

The radial profile of our MW satellite background is given by the $\Lambda$CDM prediction of the number density of subhaloes around MW-like host haloes \citep{springel08}, given by an Einasto parameters $\alpha=0.68$ and $r_{-2}=0.81R_{200}$, where $R_{200}$ is the virial radius of our MW model.
We call this (un-normalised) profile $\nu_\mathrm{MW}(r)$.
In Section~\ref{ssec:shortcomings} we will discuss the effect of loosening this assumption on the radial profile.

We parametrize the normalisation of the MW background as a fraction of the value required to reproduce the observed number of dwarfs discovered in SDSS Data Release 5 \citep[DR5,][]{adelman08} with luminosities in the range $-7<M_V<-1$.
These are shown as black points in Fig.~\ref{fig:mwmod} \citep[taken from][]{mcconnachie12}.
To determine the required normalisation we must take into account the detectability threshold of DR5, which can be approximated as a limiting luminosity as a function of distance \citep{koposov09},
\begin{equation}
\label{eqn:dr5}
M_{V,\mathrm{lim}}^\mathrm{DR5}(D_\mathrm{\sun}) = (1.1 - \log_{10} (D_\mathrm{\sun}/ \mathrm{kpc})) / 0.228,
\end{equation}
shown as a black dashed line in the top left panel of Fig.~\ref{fig:mwmod}.

To determine what fraction of dwarfs are observable subject to this criterion, we must assume a luminosity function of dwarf galaxies $L(M_V)$, i.e.
\begin{equation}
L(M_V) \propto \frac{\mathrm{d}n}{\mathrm{d}m_\mathrm{sub}} \; \frac{\mathrm{d}m_\mathrm{sub}}{\mathrm{d}m_*} \; \frac{\mathrm{d}m_*}{\mathrm{d}L} \; \frac{\mathrm{d}L}{\mathrm{d}M_V}.
\end{equation}
The first term of the right-hand side is given by the subhalo mass function, $N(M>m_\mathrm{sub})\propto m_\mathrm{sub}^{-0.9}$ \citep[e.g.][]{gao_etal_2004,springel08}.
For the second term we assume $m_\mathrm{*}\propto m_\mathrm{sub}^\gamma$, trying two values $\gamma=1.9$ \citep[][from abundance matching]{garrison14} and $\gamma=3$ \citep[][from fits to MW dwarfs]{koposov09}.
Inserting a constant stellar mass to light ratio and the standard definition of the magnitude scale into the third and fourth terms, and introducing a normalising constant $K_\gamma$ gives
\begin{equation}
L(M_V) = K_\gamma 10^{0.36 M_V / \gamma}.
\label{eqn:luminosity_func}
\end{equation}
where $K_\gamma$ is chosen to ensure that $L$ integrates to 1 in the range $-7<M_V<-1$.
The median luminosities for our two luminosity functions are $M_V=-2.9$ for $\gamma=3$ and $M_V=-2.4$ for $\gamma=1.9$, i.e. $\gamma=3$ predicts more bright dwarfs.

The number of observable dwarfs within a distance $D_\mathrm{\sun}$ is then given by the integral of the number density over the DR5 survey volume, $V$, within that distance, where we integrate the luminosity function down the faintest observable luminosity in our range, i.e. $m_v=\min\{-1, M_{V,\mathrm{lim}}^\mathrm{DR5}(D_\mathrm{\sun})\}$.
Introducing the normalisation constant $N_\mathrm{DR5}$, the expected cumulative number count is given by 
\begin{align}
N_\mathrm{exp} (<D_\mathrm{\sun}) = N_\mathrm{DR5} \int\displaylimits_{V} \nu_\mathrm{MW}(r) \int\displaylimits_{-7}^{m_v} L(M_V) \; \mathrm{d}M_V \; \mathrm{d}V.
\label{eqn:obsmod}
\end{align}
For each of our MW models and luminosity functions, we solve for the value of $N_\mathrm{DR5}$ which reproduces the observed number of dwarfs (11) in DR5 within 300 kpc.
The bottom left panel of Fig.~\ref{fig:mwmod} shows - for the luminosity function with $\gamma=1.9$ and virial radius $R_{200}=210$ kpc of our Medium MW model - our expected cumulative distribution $N_\mathrm{exp}$ of dwarfs observable in DR5 (dashed black line), alongside the observed distribution (grey histogram).
We also show the expected number count of all dwarfs over the whole sky (solid black line), i.e.
\begin{equation}
N_\mathrm{4\pi} (<D_\mathrm{\sun}) = N_\mathrm{DR5} \int\displaylimits_{V} \nu_\mathrm{MW}(r) \; \mathrm{d}V,
\end{equation}
where $V$ is now the volume within distance $D_\mathrm{\sun}$ over the whole sky.
For this choice of parameters, we predict 280 dwarfs within 300 kpc in the range $-7<M_V<-1$.

We will vary the background normalisation about $N_\mathrm{DR5}$ to account for possible anisotropy in the MW satellite distribution.
Such anisotropy is already known to exist.
We demonstrate this in Fig.~\ref{fig:mwmod} with the addition of dwarf galaxies discovered in SDSS data releases subsequent to DR5 \citep[][blue points]{grillmair09,belokurov09,belokurov10}, and the PanSTARRS1 $3\pi$ survey \citep[PS3$\pi$][cyan points]{laevens15a,laevens15b}.
The new area covered by these surveys, excluding a $20^{\circ}$ region either side of the Galactic disc where detection may be more difficult, exceeds the DR5 area by a factor of 1.6, but the number of new dwarfs discovered is a factor of 2 smaller, giving a density less than a third of that found in DR5.
We further highlight this by updating the bottom left panel with these later discoveries: the new observed number count (blue histogram), and our model distribution $N_\mathrm{exp}$ (dashed blue line) with a normalisation reduced to $1.98\times10^{-3} = 0.63 N_\mathrm{DR5}$ in order to match the observed number count within 300 kpc.
We then plot the expected number count of dwarfs over the whole sky given this normalisation (solid blue line).
The dearth of satellites in these subsequent surveys reduces our prediction of the total number of MW satellites within 300 kpc from 280 to 180.
This reduction of 100 predicted satellites is emphasised by an arrow in the bottom left panel of Fig.~\ref{fig:mwmod}.

\subsubsection{Gaussian Mixture Model}

We represent the 3D spatial probability density function (pdf) of the Magellanic distributions using a Gaussian Mixture Model (GMM) in order to smooth the discrete simulation output.
A GMM is a non-parametric density estimator consisting of a weighted sum of Gaussian components, where the weights and Gaussian parameters are found by performing a maximum-likelihood fit to the particle positions output from our simulations.

For a sample of our simulations we find the optimal number of GMM components, defined as the number which minimises the Akaike information criterion.
Each component was allowed an unconstrained covariance matrix
This optimal number was found to be in the range 7-15, with the exact solution dependent on a variety of simulation parameters, e.g. LMC satellite distributions which have undergone 1 MW crossing preferred fewer than 10 components, while those with multiple crossings preferred more than 10.
Rather than performing the costly task of finding the optimal number for each particular simulation, we chose to perform our analysis using 10 components for all of our simulated distributions.

For each combination of mass models we perform a full expectation-maximization fit for one simulation seeded from a particular set of MC kinematics, then use the resulting GMM parameters to initialise fits for the remaining simulations which are completed using 30 expectation-maximization steps.
A visual comparison of the samples drawn from the mixture models with the simulation output demonstrates that this fitting process is sufficient to capture the differences in distributions as resolved by our simulations.

\subsubsection{Likelihood}
\label{ssec:likelihood}

We model the observed satellite distribution with three components: MW, LMC and SMC.
The background MW model has a radial number density profile, 
\begin{equation}
n_\mathrm{MW}(r) = f_\mathrm{MW} N_\mathrm{DR5} \nu_\mathrm{MW} (r) ,
\end{equation}
where $r$ is Galactocentric distance and $f_\mathrm{MW}$ is a normalisation in units of the SDSS-DR5 satellite density.
The MCs have pdfs, $f_\mathrm{LMC}(\textbf{x},\boldsymbol\Theta_\mathrm{s})$ and $f_\mathrm{SMC}(\textbf{x},\boldsymbol\Theta_\mathrm{s})$, estimated via Gaussian Mixture Model, which are functions of Galactocentric Cartesian satellite position $\mathbf{x}$ and are dependent on simulation parameters $\boldsymbol\Theta_\mathrm{s}=(\mathbf{M},\boldsymbol\mu)$ where
\begin{align}
\mathbf{M} &= (M_\mathrm{MW},M_\mathrm{LMC},M_\mathrm{SMC}),\\
\boldsymbol\mu &= (\mu^\mathrm{LMC}_\mathrm{W},\mu^\mathrm{LMC}_\mathrm{N},\mu^\mathrm{SMC}_\mathrm{W},\mu^\mathrm{LMC}_\mathrm{N}),
\end{align}
i.e. the galaxy mass models and the MC proper motions, respectively.
The pdfs further depend on whether the tracer particles are initialised with either the \textit{cored} or \textit{cusped} distribution.
Introducing the parameters $N_\mathrm{LMC}$ and $ N_\mathrm{SMC}$, which we interpret as the total number of satellite galaxies which have ever belonged to the LMC or SMC, our satellite number density model, parametrized by $\boldsymbol\Theta=(\mathbf{M},\boldsymbol\mu,f_\mathrm{MW},N_\mathrm{LMC},N_\mathrm{SMC})$, is given by
\begin{equation}
F(\mathbf{x},\boldsymbol\Theta) = n_\mathrm{MW}(|\textbf{x}|) + N_\mathrm{LMC}f_\mathrm{LMC}(\textbf{x},\boldsymbol\Theta_\mathrm{s}) + N_\mathrm{SMC}f_\mathrm{SMC}(\textbf{x},\boldsymbol\Theta_\mathrm{s}).
\label{eqn:totmod}
\end{equation}

Given a selection function $S$, a model $f$ and data $x_i$, the likelihood function $\mathcal{L}$ is generically given by \citep{richardson11}
\begin{equation}
\log \mathcal{L} = - \int_S f(x) \; \mathrm{d}S + \sum_{i} \log f(x_i),
\label{eqn:likelihood_basic}
\end{equation}
where the first term is a normalising integral and the second is a summation over data points.
Our selection function is the volume $V$ of the DES footprint within $D_\mathrm{\sun}<300$ kpc where at each distance we consider only dwarfs brighter than the limiting observable luminosity $m_v=\min\{-1, M_{V,\mathrm{lim}}^\mathrm{DES}(D_\mathrm{\sun})\}$, where $M_{V,\mathrm{lim}}^\mathrm{DES}$ given by Equation~(\ref{eqn:DES_selection}), up to maximum luminosity $M_V=-7$.
To incorporate this luminosity dependent selection function into our model we must assume a luminosity function $L(M_V)$, hence our full model is the product $F(\mathbf{x},\boldsymbol\Theta) L(M_V)$.
As in Section~\ref{ssec:mwmod}, we try two luminosity functions, given by Equation~(\ref{eqn:luminosity_func}) with indices $\gamma=1.9$ or $\gamma=3$, appending this nuisance parameter to $\boldsymbol\Theta$.
We assume that the shape of the MW and MC luminosity functions are the same.
Our data, $d_\mathrm{DES} = (\mathbf{x},M_V)$, consist of the Galactocentric Cartesian positions and luminosities of the 14 satellites in Table~\ref{tab:sats}.
Inserting all of these terms into Equation~(\ref{eqn:likelihood_basic}), our likelihood function is given by
\begin{align}
\label{eqn:likelihood_pos}
\log \mathcal{L}(d_\mathrm{DES}|\boldsymbol\Theta) &= - \int_V F(\mathbf{x},\boldsymbol\Theta) \int_{-7}^{m_v} L(M_V) \;\mathrm{d}M_V \;\mathrm{d}V \nonumber \\
	& + \sum_{i=1}^{14} \log  \left( F(\mathbf{x}_i,\boldsymbol\Theta) L(M_{V,i}) \right).
\end{align}

\subsubsection{Prior PDF}

\begin{figure}
\includegraphics[width=\columnwidth]{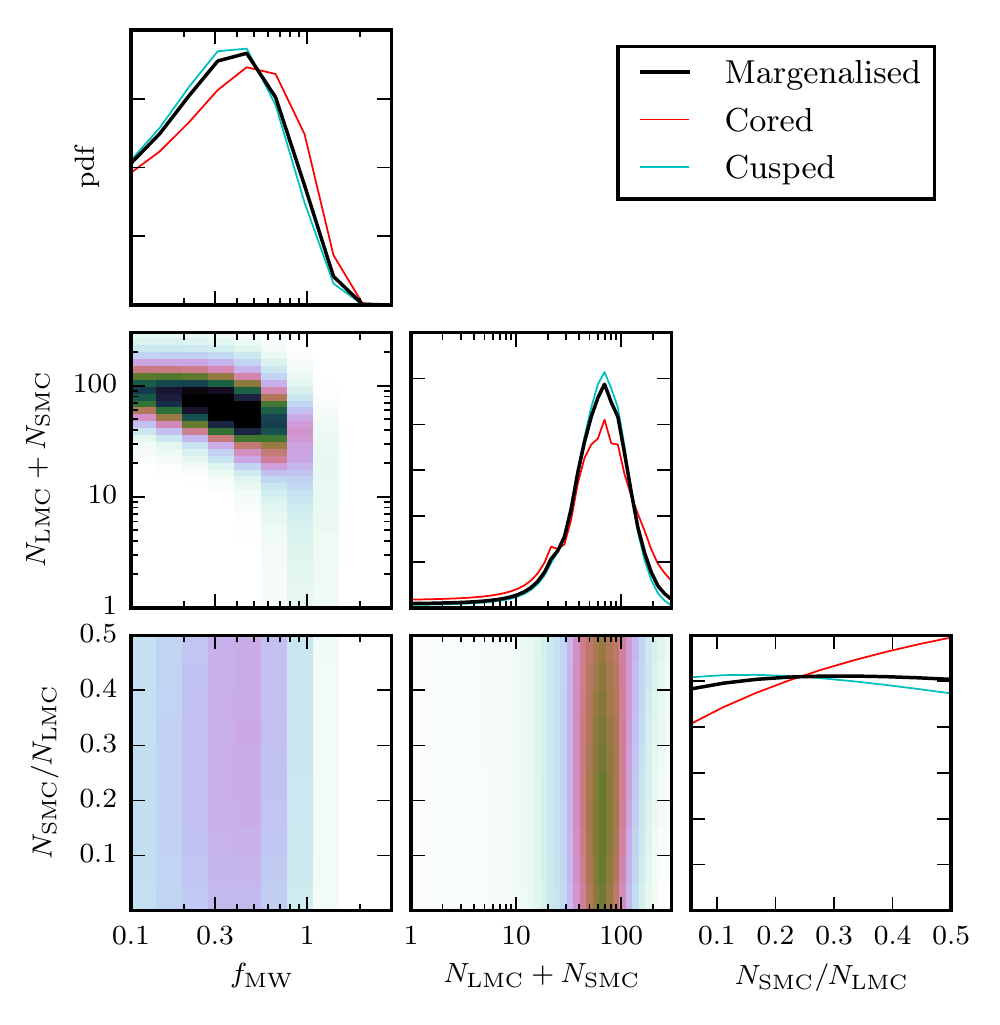}
\caption{
Triangle plot showing marginalised 1D and 2D posterior pdfs for the normalisation parameters of our model.
The parameters in columns, left to right, are the normalisation of the MW background, the total number of Magellanic satellites, and the fraction of SMC to LMC satellites.
1D posteriors show the mean pdfs in black and the dependence on the results on the initial density assumed for the satellite populations.
}
\label{fig:N_posterior}
\end{figure}

We now define the prior probability distribution on the model
parameters, $P(\boldsymbol\Theta)$.  We take a uniform prior over our
suite of simulations, in effect imposing a uniform prior over galaxy
mass models (see Table~\ref{tab:galaxy_models}) and a Gaussian prior
over proper motion measurements.
Our prior on the normalisation on
the MW background $f_\mathrm{MW}$ is uniform in log in the range
$[0.1,3]$, where $f_\mathrm{MW}=1$ gives a satellite density
comparable to that found in SDSS-DR5.

The prior on the number of satellite galaxies of the LMC is given by
\begin{equation}
\label{eqn:newprior}
P(N_\mathrm{LMC}|M_\mathrm{LMC}) \propto
\begin{cases}
	\log N_{LMC} \; &\mathrm{if} \; N_\mathrm{LMC} \leq 10 \left(\frac{M_\mathrm{LMC}}{10^{10}M_\mathrm{\sun}}\right)\\
	0 \; &\mathrm{otherwise}
\end{cases}
\end{equation}
and likewise for the SMC.
This prior is uniform in the log of number of satellite galaxies between zero and an upper limit which is dependent on the galaxy mass, allowing our lightest $2\times10^{10}M_\mathrm{\sun}$ LMC to have up to 20 satellites, whereas our most massive $3\times10^{11}M_\mathrm{\sun}$ LMC may have up to 300 satellites.
We justify this scaling of the feasible number of satellites with host mass since there is an approximately linear scaling between the number of subhaloes of a host halo and the mass of the host observed in $\Lambda$CDM simulations \citep[e.g.][]{zentner05}.

As an order-of-magnitude justification of our upper limit on satellite number, we consider the scaling between the MW and our lightest LMC.
Figure~\ref{fig:mwmod} shows that the expected number of satellites within $300$ kpc of the MW is around 200 if we assume a density of satellite galaxies over the entire sky similar to that found in the SDSS and PS$3\pi$ surveys.
Our prior allows allow our lightest LMC to have as many as 20 satellites: a factor of 10 fewer than estimated for the MW despite being a factor 50 less massive, which we consider to be a reasonably broad prior given the uncertainties in populating subhaloes with satellite galaxies.

\subsection{Results}

We now examine the posterior probability distributions of various parameters of interest.
These are calculated by evaluating the likelihood function on a grid, then using Bayes' theorem to calculate posterior probabilities,
\begin{equation}
P(\boldsymbol\Theta | d_\mathrm{DES}) = \frac{\mathcal{L}(d_\mathrm{DES}|\boldsymbol\Theta) P(\boldsymbol\Theta)}{\int_{\boldsymbol\Theta} \mathcal{L}(d_\mathrm{DES}|\boldsymbol\Theta) P(\boldsymbol\Theta) \; \mathrm{d}{\boldsymbol\Theta}},
\label{eqn:bayes}
\end{equation}
then marginalising as appropriate.

Figure~\ref{fig:N_posterior} shows the marginalised 1D and 2D posterior pdfs for three of our model parameters: the normalisation of the MW component, the total number of Magellanic satellites, and the fraction of SMC to LMC satellites.
In the 1D posterior pdfs the black lines show the result marginalised over two systematic uncertainties tested in our model, i.e. the initial density profile of MC satellites, and the luminosity function used.

We find that the results are insensitive to the choice of luminosity function.
We show the dependence of the results on our choice of initial density profile using the coloured lines in Fig.~\ref{fig:N_posterior}.
These show the marginalised posteriors when we use either the cusped (cyan) or cored (red) profiles.
We see that the fully marginalised result more closely follow the results derived from the cusped profile.
This is because, with 9 of our dozen satellites lying within 50 kpc of the LMC, our fits generically prefer the more centrally concentrated initial density profile.
Having only simulated two density slopes we refrain from further discussion of the significance of this result.
We discuss various other parameters, marginalised over these systematic uncertainties, below.

\subsubsection{The Number of Magellanic Satellites}
\label{sssec:nmagsats}

We constrain the total number of Magellanic satellites to be $70^{+30}_{-40}$, which are the mode and 68\% confidence intervals of the pdf shown in the central panel of Fig.~\ref{fig:N_posterior}.
We stress that this is not an estimate of the number we predict to be still bound to the MCs today, but refers to the total number of satellite galaxies which have evolved in the virial radii of either MC prior to their accretion onto the MW.
Many of these predicted satellites will be in tails of tidally stripped debris stretching far beyond the current extent of the matter bound to the MCs.

The bottom right panel of of Fig.~\ref{fig:N_posterior} shows that we are unable to constrain the fraction of SMC to LMC satellites.
This is in agreement with \citet{bechtol15b} (henceforth DES15).

\begin{figure}
\includegraphics[width=\columnwidth]{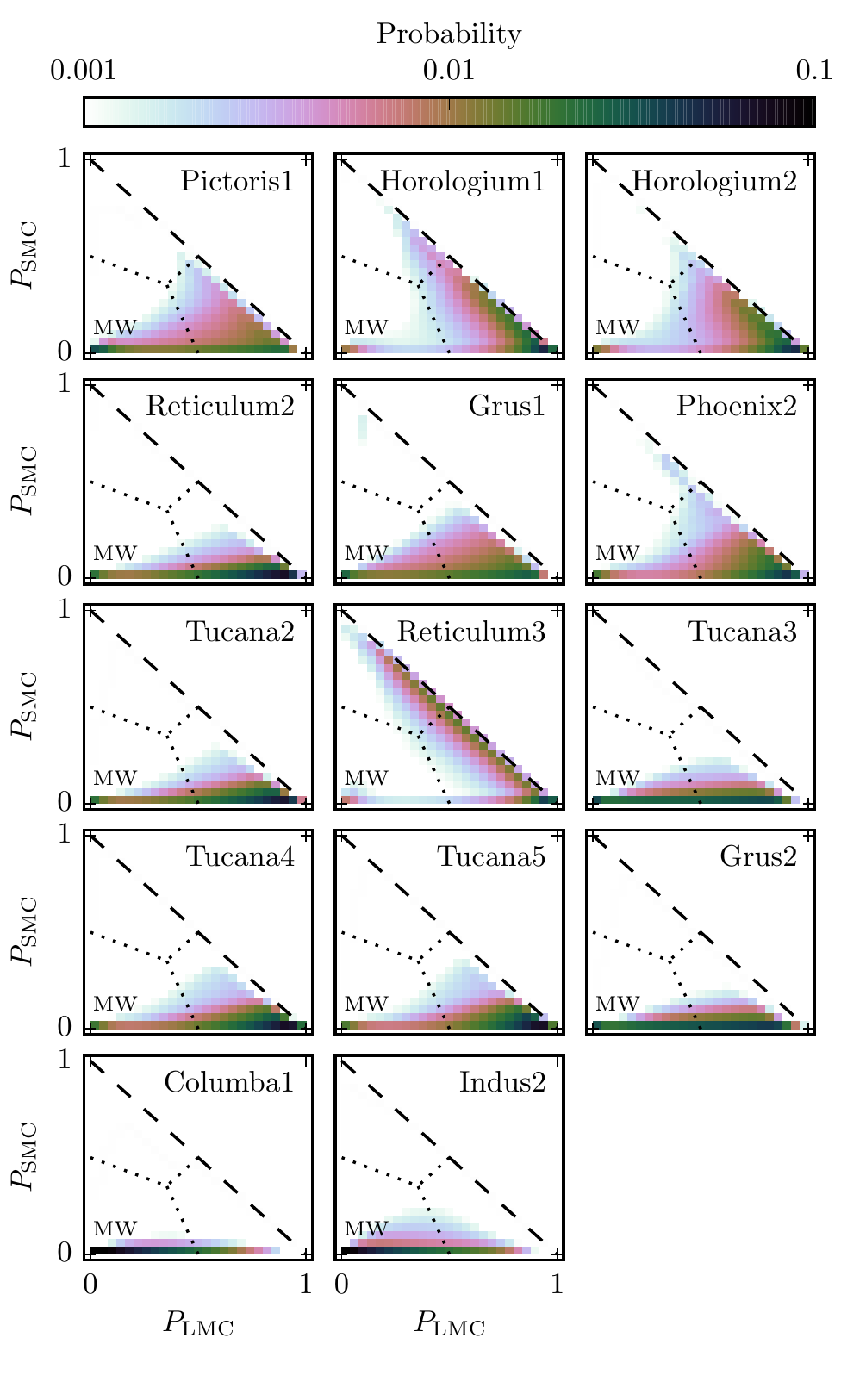}
\caption{
2D satellite membership posterior pdfs, labelled per panel.
The bottom-left/bottom-right/top-left corner of each panel imply certain membership to the MW/LMC/SMC model component respectively.
Colours show the logarithmically scaled posterior pdf in the intervening space.
Dotted lines demarcate regions where a satellite is more likely a member of one component than any other.
}
\label{fig:satprob}
\end{figure}

\subsubsection{The MW background}
\label{sssec:mwbg}

\begin{figure*}
\includegraphics[width=\textwidth]{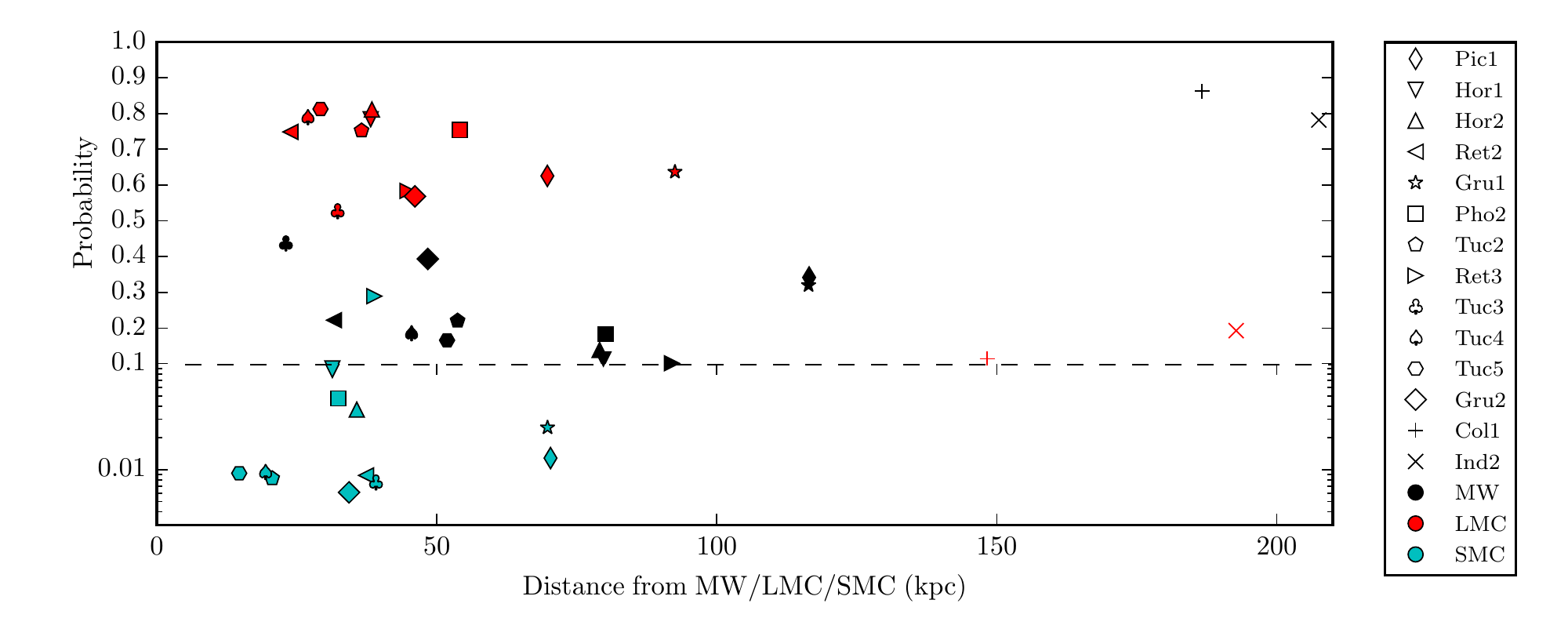}
\caption{
Probability of the observed satellites - as shown in the legend - belonging to the MW (\textit{black}), LMC (\textit{red}) or SMC (\textit{cyan}) components of our model distribution, as a function of distance from the relevant host. While there are eight satellites more likely to have originated in the LMC than the MW, there is only one satellite, Reticulum 3, with a modest probability of originating in the SMC.
}
\label{fig:satprob_r}
\end{figure*}

The left column panel of Fig.~\ref{fig:N_posterior} shows a preference for a low MW background.
This suggests that most of the DES satellites are unlikely to be drawn from an isotropic MW population.
This conclusion was previously reached in \citetalias{bechtol15b} where a simple Kolmogorov-Smirnov test significantly rejected this hypothesis.
This can be seen by eye in the right panel of Fig.~\ref{fig:mwmod}, where the on-sky positions of the DES satellites cluster around the MCs.

Quantitatively, our 68\% (95\%) confidence limits on the MW background are $f_\mathrm{MW}<0.40(0.87)$, in units of the average satellite density in SDSS-DR5 (discussed in Section~\ref{ssec:mwmod}).
We convert this into a constraint on the number of DES satellites belonging to the MW using the \citetalias{bechtol15b} estimate that roughly half of the DES detections would have been detected in SDSS.
Accounting for this extra factor of two and the DES sky coverage in Equation~(\ref{eqn:obsmod}), we arrive at a 68\% (95\%) confidence limit that fewer than 4 (8) of the 14 DES satellites shown in Fig.~\ref{fig:mwmod} belong to an isotropic MW background.

Our constraint that $f_\mathrm{MW}<1$ further implies that, once likely Magellanic satellites are excluded, the underlying MW satellite population in DES is underdense compared to SDSS-DR5.
The degree of anisotropy is weaker than that inferred from comparison between SDSS-DR5 and DR10 combined with PS3$\pi$ (as discussed in Section~\ref{ssec:mwmod}) and, of course, presupposes that none of the SDSS-DR5 satellites themselves have a Magellanic origin.

\subsubsection{Magellanic Probability of the DES satellites}
\label{sssec:magprob}

We now calculate the probability of each observed satellite belonging to either the LMC or SMC component of our model.
For model parameters, $\boldsymbol\Theta$, the probability of a satellite at position $\mathbf{x}$ belonging to the LMC component is given by
\begin{equation}
p_L(\mathbf{x}, \boldsymbol\Theta) = \frac{N_\mathrm{LMC} f_\mathrm{LMC}(\mathbf{x},\boldsymbol\Theta_\mathrm{s}) }{F(\mathbf{x},\boldsymbol\Theta)},
\end{equation}
where the total model $F(\mathbf{x},\boldsymbol\Theta)$ is defined in Equation~(\ref{eqn:totmod}).
The SMC membership probability $p_S(\mathbf{x}, \boldsymbol\Theta)$ is defined likewise.
From our simulations, we calculate the probability $P(p_L, p_S|\mathbf{x},\boldsymbol\Theta)$ that a satellite at $\mathbf{x}$ has membership probabilites $p_L$ and $p_S$ for model parameters $\boldsymbol\Theta$.
The 2D posterior pdf for a satellite at $\mathbf{x}$ to have membership probabilities $p_L$ and $p_S$, marginalised over $\boldsymbol\Theta$, is then given by
\begin{align}
P(p_L,p_S|\mathbf{x},&d_\mathrm{DES}) = \\
&\frac{1}{(*)}\int\limits_{\boldsymbol\Theta} \mathcal{L}(d_\mathrm{DES}|\boldsymbol\Theta) P(p_L, p_S|\mathbf{x},\boldsymbol\Theta) P(\boldsymbol\Theta) \; \mathrm{d}\boldsymbol\Theta , \nonumber
\end{align}
where $(*)$ is given by the denominator of Equation~(\ref{eqn:bayes}).

Figure~\ref{fig:satprob} shows the 2D membership probability pdfs calculated at the position of all 14 DES satellites.
Each panel represents the pdf for an individual satellite.
We discuss three illustrative examples:
\begin{itemize}
\item Columba 1's pdf peaks at $P_\mathrm{MW}=1$ with a little
  probability stretching towards LMC membership. Thus, we conjecture that
  Columba 1 is likely a MW satellite.
\item Tucana 5's pdf peaks at $P_\mathrm{LMC}=1$, with a smaller peak at $P_\mathrm{MW}=1$, and some probability stretching away from the $P_\mathrm{MW}=0$ axis. It is prudent to suppose that Tucana 5 could well be stripped from the LMC.
\item Reticulum 3's looks similar to Tucana 5 but has a prominent streak along $P_\mathrm{MW}=0$, stretching well into the corner of SMC membership. Therefore, there is some tantalizing evidence the Ret 3 could have been part of the SMC.
\end{itemize}

The information in Fig.~\ref{fig:satprob} is summarised in
Fig.~\ref{fig:satprob_r} where, for each DES satellite, we sum the
probability in the loci where it is more likely a member of the MW,
LMC or SMC than either other component (demarcated with dotted lines
in Fig.~\ref{fig:satprob}) and plot this as a function of distance
from the relevant host. Note a large number of satellites (likely of
the LMC origin) for which the red filled symbol (the probability of
belonging to the LMC) lies above the corresponding black filled symbol
(the MW membership). For such pairs, the distances from the LMC and
the MW are sorted in an obvious manner: $D_{\rm LMC} < D_{\rm MW}$.

Table~\ref{tab:plmc} summarises LMC membership probabilities for the DES satellites, which we split into three bins comprising 7 likely LMC members, 5 possible members and 2 unlikely members.
Comparing with table 1 of \citet{deason15}, we typically assign a slightly greater probability of LMC membership ($\Delta p_\mathrm{LMC}\sim0.2$).
We attribute this to the fact that the relative normalisation of the MW background is a free parameter in our model, whereas in \citet{deason15} it is fixed by their simulations.

\begin{table}
	\renewcommand{\arraystretch}{1.0}
	\centering
	\caption{DES satellites split by LMC membership probabilities.}
	\label{tab:plmc}
	\begin{tabular}{cc}
	\hline\hline
	LMC membership 	& Satellites 	\\ \hline

	\multirow{2}{*}{$p_\mathrm{LMC}>0.7$}	& Ret2, Tuc2, Tuc4, Tuc5	\\
											& Hor1, Hor2, Pho2			\\ \hline
	$0.5<p_\mathrm{LMC}<0.7$				& Ret3, Tuc3, Gru1, Gru2, Pic1	\\ \hline
	$p_\mathrm{LMC}<0.2$					& Col1, Ind2			\\ \hline
	\end{tabular}
\end{table}

\subsubsection{Influence of the SMC?}

Despite not being able to attribute an SMC origin to any of the DES satellites with a high probability, we can also ask whether its gravitational force has been influential for the orbital history of any of the DES satellites assuming they have an origin in the LMC satellite population.
As a first pass in answering this, we perform a further grid of simulations with no SMC, i.e. we simulate the LMC satellite population in the potential of the MW and LMC only.
Then, over the entire grid of simulations, we calculate the likelihood of the data, $d_\mathrm{DES}$, in the two component MW+LMC model, given by Equation~(\ref{eqn:totmod}) excluding the SMC component.
Doing this, we discount the possibility that the SMC has its own satellites, but it still affects results through its gravitational influence on LMC satellites.

Starting from a flat prior over the values the mass ratio $M_\mathrm{LMC}/M_\mathrm{SMC}=(3,5,10,\infty)$ (where $\infty$ means no SMC) gives a posterior pdf $P_\mathrm{posterior}=(0.29,0.26,0.25,0.19)$.
The result does not deviate greatly from the prior, however we slightly prefer every instance of the LMC populations which have experienced a massive SMC to those which have not.
This very tentatively suggests that the SMC has had some influence over at least one of the DES satellites.
We defer further investigation of this to later work.

\subsubsection{Number of MW-LMC Pericenters}

We can use the positions of the DES satellites to try to determine the number of pericenters undergone between the MW and LMC in the last 11.5 Gyr.
Our prior on this quantity, $P(N_\mathrm{peri}=n)$, comes from counting the fraction of MW-LMC orbits calculated, as described in Section~\ref{ssec:orbint}, which have exactly $n$ pericenters.
Our posterior, given the observed positions of the DES satellites, is then
\begin{align}
P(N_\mathrm{peri}&=n|d_\mathrm{DES}) =\\ &\frac{1}{(*)} \int\limits_{\boldsymbol\Theta}  \mathcal{L}(d_\mathrm{DES}|\boldsymbol\Theta) P(N_\mathrm{peri}=n|\boldsymbol\Theta) P(\boldsymbol\Theta) \; \mathrm{d}\boldsymbol\Theta , \nonumber
\end{align}
where the normalisation $(*)$ is given by the denominator of Equation~(\ref{eqn:bayes}).

\begin{table}
	\centering
	\caption{Prior and posterior probability distributions on the number of pericenters undergone between the MW and LMC in the last 11.5 Gyr.}
	\label{tab:nperi}
	\begin{tabular}{ccccc}
	\hline\hline
	\# MW-LMC pericenters	& $P_\mathrm{prior}$ & $P_\mathrm{posterior}$ \\ \hline
	1						& 0.64	& 0.90		\\
	2						& 0.29	& 0.10		\\
	>2						& 0.07	& $<10^{-3}$		\\ \hline
	\end{tabular}
\end{table}

Table~\ref{tab:nperi} shows the results.
With the mass models we have used for the MW and LMC, and the measurements adopted for the LMC velocity, 64\% of our orbit integrations result in a ``first-passage" solution, where the LMC has only recently (<0.1 Gyr ago) completed is first pericenter about the MW.
Updated with our knowledge of the positions of the DES satellites, the probability of a first-passage solution increases to 0.90.
This is because at each pericenter the LMC undergoes an enhanced bout of tidal stripping, making it less likely to retain the localised concentration of satellites that has been observed.
This echoes the recent result of \citet{deason15}, that the DES satellites suggest a recent accretion of the LMC onto the MW, as well as previous works which suggested that the LMC is on its first infall \citep[e.g.][]{besla07}.

Figure~\ref{fig:nperi} shows that the number of MW-LMC pericenters is primarily controlled by the mass of the MW.
Our preference for ``first-passage" solutions therefore converts into a preference for a low mass MW.
For a flat prior over the three values, $M_{200}^\mathrm{MW}=(50,100,200)\times10^{11}M_\mathrm{\sun}$ gives $P_\mathrm{posterior}=(0.65,0.28,0.07)$.

\subsection{Maximum Likelihood 3D Model}
\label{ssec:mlmod}

\begin{figure*}
\includegraphics[width=\textwidth]{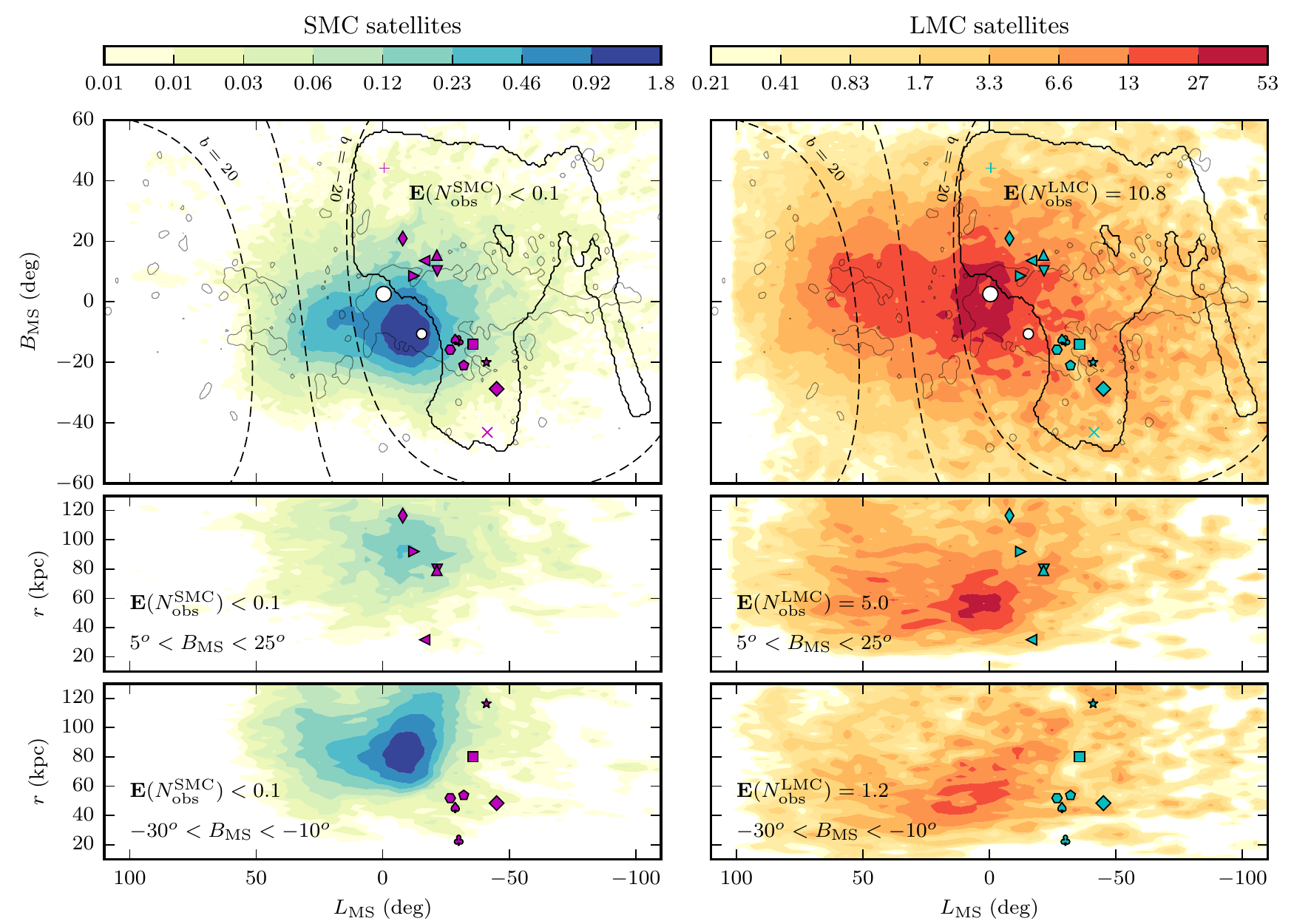}
\caption{
Maximum likelihood model of the satellites of the SMC (left) and LMC (right).
\textbf{Top row}: on-sky $(L_\mathrm{MS},B_\mathrm{MS})$ projection of (i) simulated satellite distribution (coloured contours), (ii) the LMC/SMC (large/small white circles), (iii) DES satellites (coloured symbols defined in Fig.~\ref{fig:satprob_r}), (iv) DES footprint (solid line), (v) distribution of HI gas (faint contours), (vi) $20^\circ$ either side of the Galactic disk (dashed lines).
\textbf{Middle/bottom rows}: distribution in $L_\mathrm{MS}$ against Galactocentric $r$ for bins $5^\circ<B_\mathrm{MS}<25^\circ$ (middle) and $-30^\circ<B_\mathrm{MS}<-10^\circ$ (bottom).
Contours step roughly in factors of 2 in projected density, with arbitrary units and colours comparable per column.
For the maximum likelihood solution of $\sim70$ Magellanic satellites, the expected number of satellites observable by DES is annotated in each panel.
}
\label{fig:bestfit}
\end{figure*}

Figure~\ref{fig:bestfit} shows the satellite distributions from our maximum likelihood model.
We show the on-sky distribution in Magellanic Stream (MS) co-ordinates $(L_\mathrm{MS},B_\mathrm{MS})$ defined in \citet{nidever08}, and in Galactocentric $r$ against $L_\mathrm{MS}$ in two bins of $B_\mathrm{MS}$, for both the LMC and SMC components.
These have been generated from simulations using a cusped initial satellite distribution, galaxy masses ($M_\mathrm{MW}, M_\mathrm{LMC}, M_\mathrm{SMC}) = (50, 12, 4) \times 10^{10} M_\mathrm{\sun}$, and combining over all MC proper motions tried.
The maximum likelihood normalisations are $f_{MW}=0.2$, $N_\mathrm{LMC}=70$ and $N_\mathrm{SMC}=1$, for a luminosity function with $\gamma=1.9$.
We convert these figures into the expected number of observed satellites by integrating our satellite density models taking into account the survey detection limit as given by Equation~(\ref{eqn:DES_selection}).
Inside the entire DES volume, this results in expected numbers of 10.8 observable LMC, 2.7 MW and fewer than 0.1 SMC satellites.

\begin{figure}
\centering
\includegraphics[]{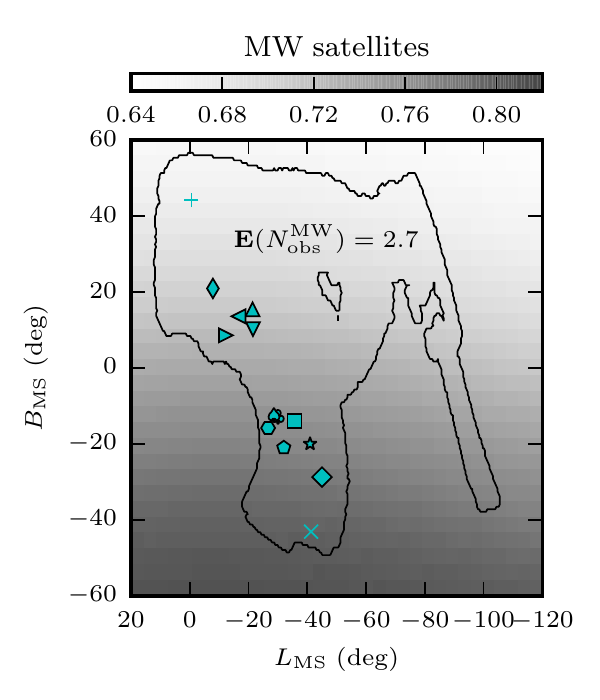}
\caption{
On-sky density model of MW dwarfs.
Symbols show DES satellites, defined in Fig.~\ref{fig:satprob_r}.
We also show the DES footprint, and label the expected number of MW satellites observable in DES.
}
\label{fig:mw_sats}
\end{figure}

Figure~\ref{fig:mw_sats} shows the on-sky density of satellites for our background MW model.
The image in $(L_\mathrm{MS},B_\mathrm{MS})$ shows the quantity
\begin{equation}
f(L_\mathrm{MS},B_\mathrm{MS}) = \int\displaylimits_{D_\mathrm{\sun}=0}^{300 \; \mathrm{kpc}} n_\mathrm{MW}(r) D_\mathrm{\sun}^2 \; \mathrm{d}D_\mathrm{\sun}
\end{equation}
where we have not included a $\cos(B_\mathrm{MS})$ volume correction for clarity.
Our MW component has a density gradient across the DES footprint, being $\sim20\%$ greater in the south west corner of the DES footprint (i.e. the bottom left corner of the plot) that the north east.
This is due to the relative proximity of the south west corner to the Galactic center.
A line of sight through this region therefore traverses more of the MW halo and has a greater probability of encountering a MW satellite.
Our model naturally accounts for this gradient, however we find it is not sufficient to account for the observed distribution of the DES satellites and we require a significant Magellanic contribution, as shown in Fig~\ref{fig:satprob_r}.

This maximum likelihood model assigns 12 of the DES satellites a probability $p_L>0.7$ of belonging to the LMC, while the more distant Columba 1 and Indus 2 are most likely to be MW satellites.
The middle left panel of Fig.~\ref{fig:bestfit}, we see that see that the distance of Reticulum 3 coincides with the peak of the distribution of SMC satellites, making it the only DES satellite with a non-negligible probability of belonging to the SMC.
All other satellites are much more likely to belong to the background MW population than to the SMC.

\begin{figure}
\includegraphics[width=\columnwidth]{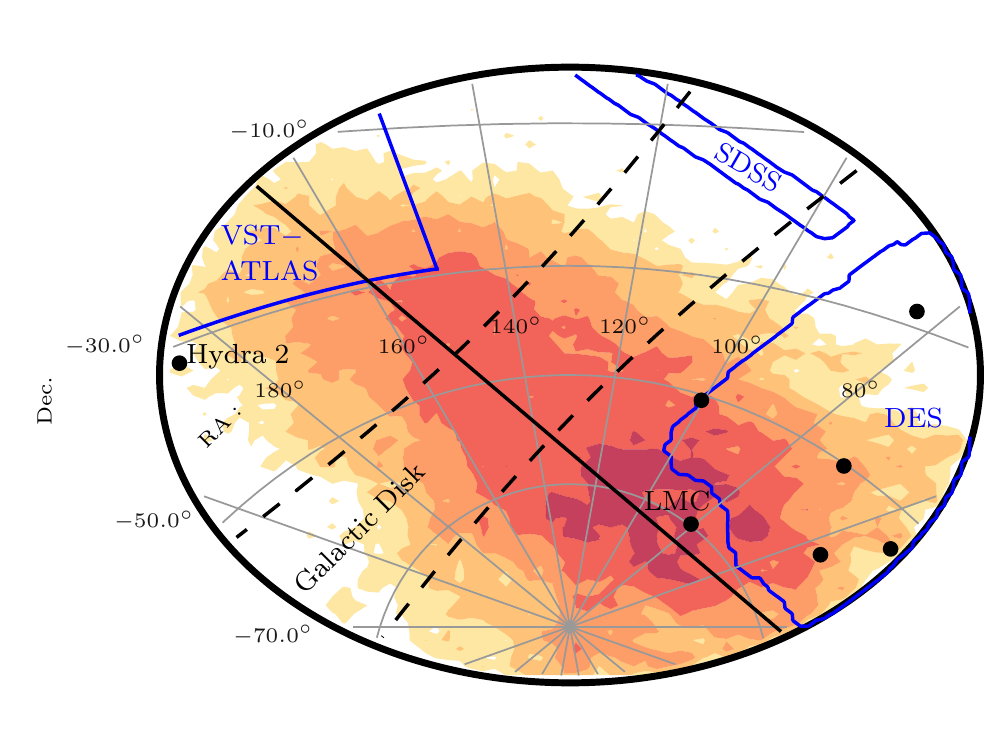}
\caption{
Leading arm of Magellanic satellites shown in equatorial coordinates in a gnomic projection centred on (RA, Dec.)=($130^\circ,-50^\circ)$.
Filled coloured contours represent the distribution of satellites from our maximum likelihood model integrated along the line of sight.
Footprints of the SDSS, DES, and ATLAS surveys are shown alongside lines of constant Galactic latitude at $b=-10^\circ$ and $b=10^\circ$.
The solid black line tracing the centre of the distribution is a segment of the great circle with pole (RA, Dec.)=($275^\circ,-41^\circ)$.
Black dots show the positions of known satellites, with the LMC and Hydra 2 labelled.
}
\label{fig:leadingarm}
\end{figure}

Most of the remainder of the predicted LMC satellites are located in a
leading arm of debris stretching out behind the Galactic disk.  This
is shown in equatorial co-ordinates in Fig.~\ref{fig:leadingarm},
along with the footprints of SDSS, DES and ATLAS \citep{shanks15}
surveys.  Much of the predicted leading arm is currently lacking
contiguous coverage, potentially making the high Galactic latitude
portion of the area shown in the figure a fruitful region for future
satellite searches.  Curiously, despite a rather patchy sky coverage,
the Survey of the Magellanic Stellar History (SMASH; PI D. Nidever)
has already serendipitously unearthed the Hydra II dwarf galaxy
\citep{martin15} at a position consistent with the LMC leading tail.
The Galactocentric distances of our simulated leading arm of
satellites peak in the range 40-80 kpc, albeit extending as far as 300
kpc.

While our model can reproduce the number of satellites discovered, there is tension regarding their distribution within the DES footprint.
The top right panel of Fig.~\ref{fig:bestfit} shows that our model predicts a concentration of satellites within ~10 kpc of the LMC, which is not present in the data.
This might be a result of our over-simplistic treatment of the tidal disruption of satellites.
The stellar disk of the LMC extends beyond 10 kpc from its centre \citep[e.g.][]{vdmarel14}, hence satellite destruction is likely to occur at distances greater than the 5 kpc zone-of-avoidance we have naively imposed in our model.
For our maximum likelihood model, the fraction of particles today within 10, 15 and 20 kpc of the LMC is 3, 8 and 14 \% respectively; accounting for destruction of satellite which have, at any time, entered within these distances would reduce our prediction of the total number of Magellanic satellites by at least these proportions.

A further failure of our model is the under-prediction of satellites in the range $-30^\circ<B_\mathrm{MS}<-10^\circ$.
Our maximum likelihood model predicts 1.2 satellites in this range, whereas 7 are observed.
Even allowing for the possibility that some of these belong to a background MW population, this suggests an over-density of satellites in a small spatial region compared to expectations from the disruption of an isotropic LMC satellite population.
Furthermore, despite their on-sky proximity to the SMC, Fig.~\ref{fig:satprob_r} shows that 5 of these 7 low $B_\mathrm{MS}$ satellites (Tucana 2-5 and Grus 2) are in fact the least likely candidates to have belonged to the SMC.
This is due to their distances ($r<60$ kpc) being incompatible with satellites stripped from the SMC halo, as seen in the bottom left panel of Fig.~\ref{fig:bestfit}.

\begin{figure}
\includegraphics[width=\columnwidth]{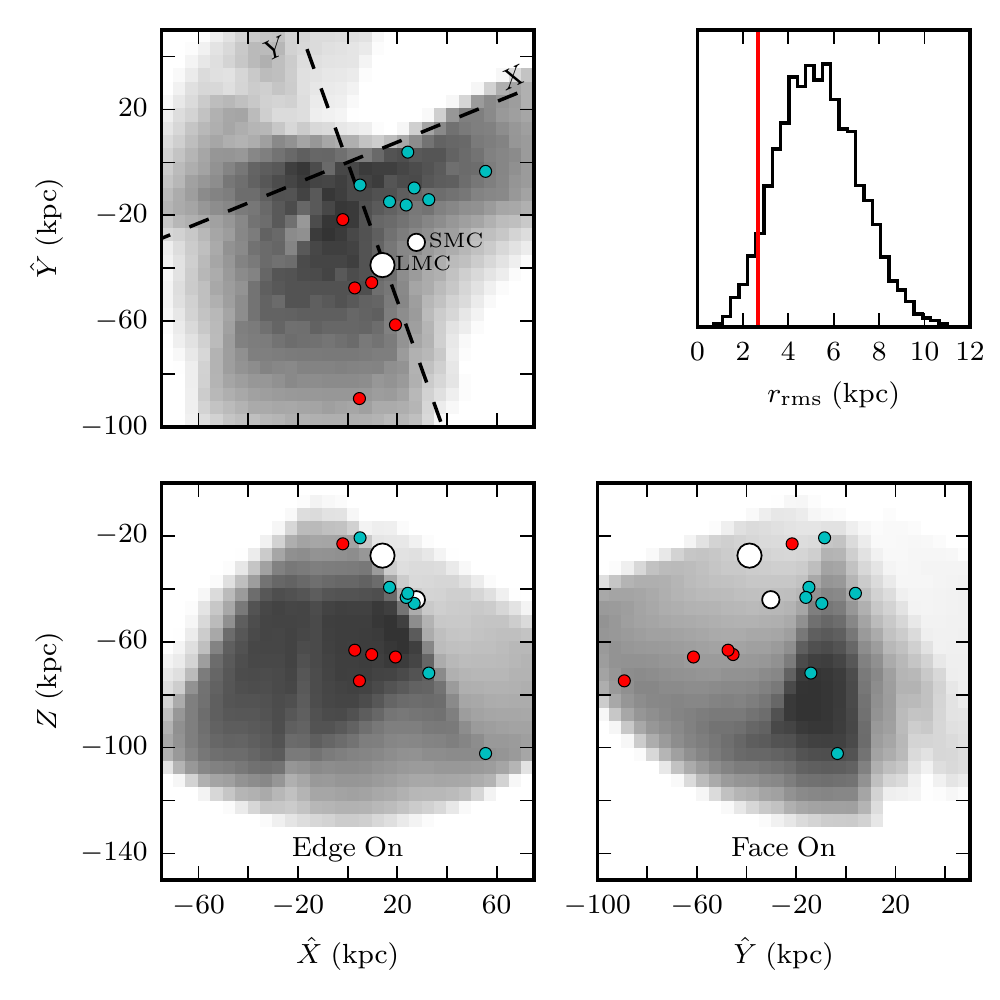}
\caption{
Distribution of $B_\mathrm{MS}>0^\circ$ satellites (red) and $B_\mathrm{MS}<0^\circ$ (cyan) in rotated Galactocentric Cartesian co-ordinates.
The LMC/SMC are shown by the large/small white circles, the projection of the volume limited by the DES footprint is shown in grey-scale.
The top right panel shows the distribution of root-mean-squared distances of groups of 7 low $B_\mathrm{MS}$ satellite analogues from our simulations, with the observed value in red.
We conclude that the observed plane has a significance of 95\%
}
\label{fig:plane}
\end{figure}

\subsection{A Plane of LMC Satellites?}

We now investigate this hint of an anisotropy in the LMC satellite population.
We investigate whether the seven DES satellites observed at $B_\mathrm{MS}<0^\circ$ reside in a coherent structure, namely a thin plane as has been observed amongst the satellites of both the MW \citep[e.g.][]{kroupa05} and M31 \citep{ibata13}.
We test $10^4$ normal vectors spaced uniformly around the unit sphere to find the plane containing the LMC which minimises the root-mean-squared (rms) distance of the seven $B_\mathrm{MS}<0^\circ$ DES satellites.
The resulting plane has an rms thickness of 2.7 kpc, with a radial extent of 90 kpc to the most distant satellite, Grus 1.
The SMC has a minimum distance of 2.0 kpc from the plane.

Figure~\ref{fig:plane} shows the satellite distribution in Galactocentric Cartesian co-ordinates $(\hat{X}, \hat{Y},Z)$, which have been rotated by $20.94^{\circ}$ about the $Z$-axis such that the best-fit plane is edge-on in the $(\hat{X},Z)$ projection.
We also show the projection of the volume limited by the DES footprint, which appears to coincide with the plane in the $(\hat{X},Z)$ projection.
This is a mere coincidence, since the low $B_\mathrm{MS}$ satellites do not lie near the footprint edges, as can be seen in Fig.~\ref{fig:bestfit}.

We test the significance of this observed satellite plane by comparing to our simulations.
We take our maximum likelihood model and consider all the particles representing LMC satellites in the box $-50^\circ<L_\mathrm{MS}<-25^\circ$, $-30^\circ<B_\mathrm{MS}<-10^\circ$, which number around $10^4$.
We then consider $10^4$ random samples of 7 of these particles, insisting that the particle furthest from the LMC has a distance within 90-100 kpc, similar to Grus 1 at 93 kpc.
We perform the same plane fitting procedure on the random samples as with the observed satellites.
The top right panel of Fig.~\ref{fig:plane} shows the distribution of the resulting rms distances.
Finding a plane at least as thin as the one observed occurs 5\% of the time when sampling from our simulations, i.e. the plane has a significance of 95\%.

This calculation does not take into the account the fact that some of the low $B_\mathrm{MS}$ satellites may not be genuine LMC satellites.
Figure~\ref{fig:satprob_r} shows that three of these (Tucana 3, Grus 1 and Grus 2) have MW membership probabilities $>0.5$.
None are likely SMC candidates.
Accounting for their uncertain origin, then, would increase the significance (and coincidental nature!) of the observed LMC satellite plane, since the distance distribution of MW satellites is broader than that of LMC satellites.

The observation of this plane is thus quite surprising.
Given a population of satellites which is initially distributed isotropically about the LMC, then evolved in the combined potentials of the MW, LMC and SMC, it is a rare occurance to find 7 satellites - selected on similar criteria to the observed ones - exhibiting the observed degree of planarity.

\section{Model for the Velocity Distribution}
\label{sec:veloc}

We now consider the velocity information contained in our model.
To recap, the velocities of our simulated satellite are composed of the orbital velocity of the LMC or SMC about the MW (or indeed the SMC about the LMC), a dispersion inherent in our simulated MC haloes, and a component induced by the potential of the MW for any satellite which has been, or is in the process of being, tidally stripped from its host.
In addition to their configuration-space positions, satellite velocity information can act as a discriminant of membership to the MCs, and trace the gravitational potential of their host at distances unprobed by other luminous tracers.

In this section we will look at the phase-space distribution of our maximum likelihood model, then derive marginalised velocity predictions for the DES satellites assuming their membership to the MCs.

\subsection{Maximum Likelihood Model for Kinematics}

\begin{figure}
\includegraphics[width=\columnwidth]{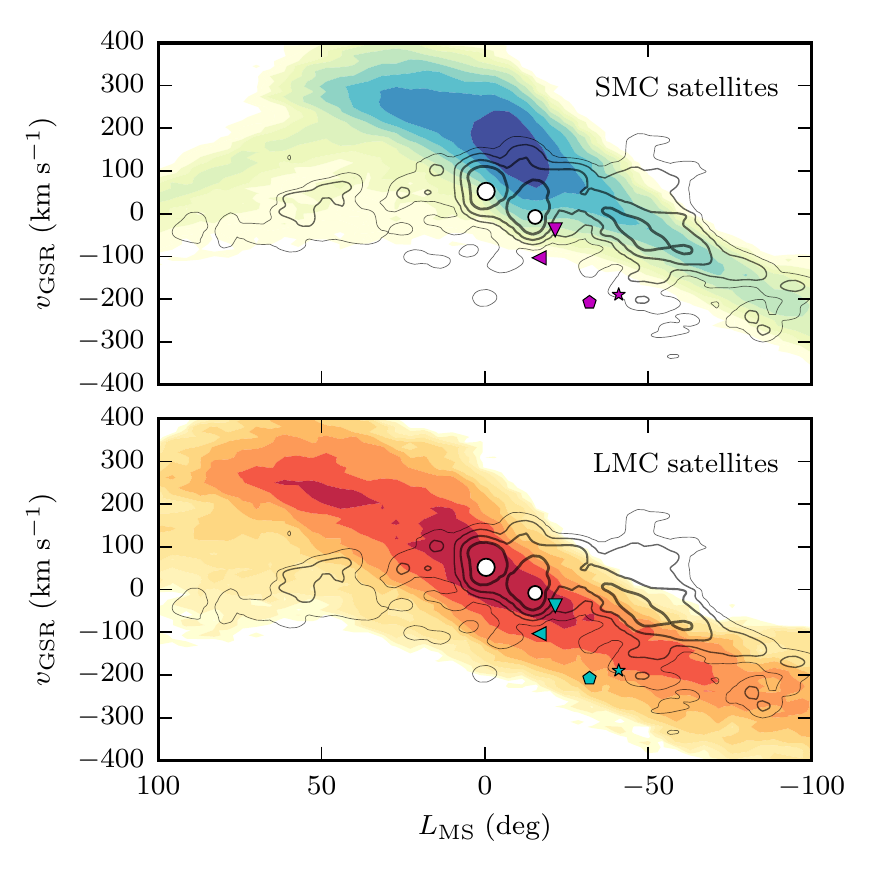}
\caption{Distributions in GSR line-of-sight velocity against $L_\mathrm{MS}$.
Filled contours represent the simulated satellites for our maximum likelihood model of SMC (top) and LMC (bottom), coloured symbols show the positions of four DES satellites (given in the legend of Fig.~\ref{fig:satprob_r}), large/small white circles show the position of the LMC/SMC, black contours show the distribution of HI gas from \protect\cite{nidever08}.
}
\label{fig:mlvelocity}
\end{figure}

Figure~\ref{fig:mlvelocity} shows a projection of the phase-space distribution of the LMC and SMC satellite populations of our maximum likelihood model defined in Section~\ref{ssec:mlmod}.
We show the distribution in $L_\mathrm{MS}$ against $v_\mathrm{GSR}$, the line-of-sight velocity in the Galactic standard of rest (GSR).
We also show the positions of the MCs and the positions of the four DES satellites with measured line-of-sight velocities which are given in the rightmost column of Table~\ref{tab:velocities}.

Viewed in this projection, all four DES satellites with measured velocities appear consistent with the simulated LMC satellite distribution.
Furthermore, Tucana 2 and Grus 1 (at $L_\mathrm{MS}\approx-40^\circ$) have more negative velocities than Horologium 1 and Reticulum 2 (at $L_\mathrm{MS}\approx-20^\circ$).
This is consistent with the velocity trend seen in the simulated trailing arm of LMC satellites, bolstering the case for their membership to this population.

\subsection{Comparison to HI gas}

Figure~\ref{fig:mlvelocity} also shows is the distribution of HI gas taken from \citet{nidever08} overplotted as black contours.
Contour levels step logarithmically, with line thickness representing density projected in the $(L_\mathrm{MS},v_\mathrm{GSR})$ plane.
This distribution comprises gas within the LMC and SMC, a bridge of gas connecting the two, the trailing Magellanic Stream (MS) at longitudes $L_\mathrm{MS}<0^\circ$, and a few complexes comprising the Leading Arm (LA) at longitudes $L_\mathrm{MS}>0^\circ$.

We see that the MS gas (at $L_\mathrm{MS}<0^\circ$) is systematically
offset from the observed LMC satellite velocites by some 100-150 km
s$^{-1}$.  Assuming a scenario where the LMC originally played host to
some or all of the MS gas \citep[as proposed in][]{nidever08}, and the
satellites (as proposed here), this offset may be the result of
ram-pressure stripping of the MS gas by the hot MW halo gas, an effect
to which the satellite orbits remain impervious.

Alternatively, this offset may be the result of the MS originating in the SMC rather than the LMC.
Indeed, the models of both \citet{besla12} and \citet{diaz12} show that tidal stripping of gas from the SMC is able to reproduce morphological and kinematic properties of the MS.
Though these models tidally strip gas from a rotating SMC disc, whereas we strip satellites from an isotropic spherical distribution, the fact that the MS and our simulated SMC satellite kinematics are broadly comparable, as shown in the top panel of Fig.~\ref{fig:mlvelocity}, is reassuring.

We also mention the recent MS production model of \citet{hammer15},
which invokes a collision between the MCs and ram-pressure stripping
to successfully reproduce many morphological and kinematic properties
of the MS.  Given the inability of ram-pressure to pull gas in front
of the MC orbits, they propose that the Leading Arm (LA) may be the
result of ``successive passages of leading, small dwarfs that have
lost their gas from ram pressure''. Fig.~\ref{fig:bestfit} shows that
the on-sky positions of the gas complexes comprising the LA are
comfortably within the scope of our simulated LMC satellite
distribution, hence we believe that the LA production scenario
described in \citet{hammer15} is well accommodated in the picture of
Magellanic group infall.  \citet{hammer15} presuppose that the
inferred time gap between the arrival to the MW of the dwarfs and the
MCs precludes their belonging to group, however this fails to consider
the possibility of very massive
($M_\mathrm{200}>10^{11}M_\mathrm{\sun}$) and extended LMC dark halo.

\subsection{Velocity predictions}

We now predict the velocities of the DES satellites under the assumption that they were members of one of the MCs, marginalised over our entire grid of models.
For each of our simulations, $\boldsymbol\Theta$, we approximate the local velocity distribution at position $\mathbf{x}$ assuming membership to the LMC population - which we call $P(v|\mathbf{x},\boldsymbol\Theta,\mathrm{LMC})$ - as a normal distribution with mean and standard deviation given by the velocities of the ten LMC satellite tracers nearest to $\mathbf{x}$, excluding any three-sigma velocity outliers.
We then marginalise over model parameters to give velocity pdfs for a satellite at $\mathbf{x}$, assuming it belongs to the LMC,
\begin{align}
P(v|\mathbf{x}&,d_\mathrm{DES},\mathrm{LMC}) =\\
&\frac{1}{(*)} \int\limits_{\boldsymbol\Theta} \mathcal{L}(d_\mathrm{DES}|\boldsymbol\Theta) P(v|\mathbf{x},\boldsymbol\Theta,\mathrm{LMC}) P(\boldsymbol\Theta)\; \mathrm{d}\boldsymbol\Theta , \nonumber
\end{align}
where $(*)$ is the denominator of Equation~(\ref{eqn:bayes}), and we have also incorporated the information provided by the positional data $d_\mathrm{DES}$.
We derive velocity prediction assuming SMC membership similarly.

\begin{table*}
	\centering
	\caption{Satellite velocity predictions and measurements.
	Predictions are given for line-of-sight velocity $v_\mathrm{hel}$ and proper motions $(\mu_W,\mu_N)$ in the solar rest-frame, assuming membership to the LMC (columns 2-4) and SMC (columns 5-7).
	We highlight the SMC prediction for the most likely candidate, Reticulum 3.
	Predictions are given as a maximum likelihood value with asymmetric errors indicating the 68\% (95\%) confidence intervals.
	Column 8 shows measured line-of-sight velocities from (1) \citet{koposov15b} (2) \citet{walker15}.
	}
	\label{tab:velocities}
	\begin{tabular}{llllllll}
	\hline\hline
					& \multicolumn{3}{c}{LMC Predictions}                       & \multicolumn{3}{c}{SMC Predictions}                       & \multicolumn{1}{c}{Observed}		\\
Name				& \multicolumn{1}{c}{$v_\mathrm{hel}$}  & \multicolumn{1}{c}{$\mu_W$}           & \multicolumn{1}{c}{$\mu_N$}           & \multicolumn{1}{c}{$v_\mathrm{hel}$}  & \multicolumn{1}{c}{$\mu_W$}           & \multicolumn{1}{c}{$\mu_N$}           & \multicolumn{1}{c}{$v_\mathrm{hel}$}	\\
					& \multicolumn{1}{c}{(km s$^{-1}$)}     & \multicolumn{1}{c}{(mas yr$^{-1}$)}   & \multicolumn{1}{c}{(mas yr$^{-1}$)}   & \multicolumn{1}{c}{(km s$^{-1}$)}     & \multicolumn{1}{c}{(mas yr$^{-1}$)}   & \multicolumn{1}{c}{(mas yr$^{-1}$)}   & \multicolumn{1}{c}{(km s$^{-1}$)}		\\
	\hline
Pic1	& $245^{+87(180)}_{-66(148)}$	& $-0.22^{+0.17(0.38)}_{-0.19(0.41)}$	& $-0.87^{+0.16(0.39)}_{-0.15(0.33)}$	& $344^{+70(151)}_{-84(596)}$	& $-0.50^{+0.36(0.64)}_{-0.17(0.62)}$	& $-0.71^{+0.17(0.72)}_{-0.16(0.41)}$	& -	\\[0.25cm] 
Hor1	& $124^{+88(204)}_{-70(157)}$	& $-0.75^{+0.25(0.57)}_{-0.27(0.59)}$	& $-1.28^{+0.20(0.45)}_{-0.24(0.51)}$	& $270^{+87(188)}_{-76(526)}$	& $-1.08^{+0.40(1.08)}_{-0.22(0.59)}$	& $-0.97^{+0.22(0.98)}_{-0.20(0.46)}$	& 112.8$^{(1)}$	\\[0.25cm] 
Hor2	& $133^{+89(202)}_{-71(158)}$	& $-0.71^{+0.26(0.58)}_{-0.27(0.60)}$	& $-1.32^{+0.20(0.45)}_{-0.24(0.51)}$	& $271^{+87(190)}_{-75(528)}$	& $-1.09^{+0.37(1.09)}_{-0.23(0.64)}$	& $-0.99^{+0.22(1.00)}_{-0.20(0.48)}$	& -	\\[0.25cm] 
Ret2	& $186^{+125(288)}_{-107(237)}$	& $-1.37^{+1.46(2.98)}_{-1.20(2.77)}$	& $-3.34^{+1.02(2.10)}_{-0.74(1.71)}$	& $362^{+147(329)}_{-194(660)}$	& $-0.00^{+-0.32(1.48)}_{-3.63(5.54)}$	& $0.02^{+-1.01(0.13)}_{-3.22(4.49)}$	& 64.7$^{(1)}$	\\[0.25cm] 
Gru1	& $-61^{+84(187)}_{-65(139)}$	& $-0.93^{+0.12(0.29)}_{-0.17(0.35)}$	& $-0.27^{+0.15(0.32)}_{-0.16(0.33)}$	& $42^{+101(297)}_{-95(332)}$	& $-0.87^{+0.12(0.86)}_{-0.22(0.58)}$	& $-0.23^{+0.14(0.31)}_{-0.20(0.64)}$	& -140.5$^{(2)}$	\\[0.25cm] 
Pho2	& $-30^{+82(193)}_{-71(157)}$	& $-1.33^{+0.19(0.43)}_{-0.26(0.54)}$	& $-0.61^{+0.25(0.55)}_{-0.27(0.58)}$	& $133^{+105(244)}_{-114(418)}$	& $-1.31^{+0.27(1.31)}_{-0.34(0.71)}$	& $-0.55^{+0.23(0.58)}_{-0.27(0.63)}$	& -	\\[0.25cm] 
Tuc2	& $-34^{+110(256)}_{-80(181)}$	& $-2.09^{+0.31(0.76)}_{-0.41(0.89)}$	& $-0.58^{+0.47(1.09)}_{-0.52(1.18)}$	& $169^{+126(272)}_{-153(486)}$	& $-0.00^{+-1.26(0.00)}_{-2.50(3.15)}$	& $0.02^{+-0.25(0.27)}_{-1.23(1.84)}$	& -129.1$^{(2)}$	\\[0.25cm] 
Ret3	& $205^{+88(196)}_{-66(148)}$	& $-0.37^{+0.22(0.49)}_{-0.24(0.52)}$	& $-1.11^{+0.16(0.39)}_{-0.20(0.41)}$	& $\mathbf{330^{+68(155)}_{-68(584)}}$	& $\mathbf{-0.56^{+0.26(0.59)}_{-0.27(0.67)}}$	& $\mathbf{-0.86^{+0.15(0.87)}_{-0.18(0.40)}}$	& -	\\[0.25cm] 
Tuc3	& $47^{+188(396)}_{-148(332)}$	& $-3.77^{+2.72(5.67)}_{-1.54(3.71)}$	& $-2.20^{+2.11(4.89)}_{-1.78(4.14)}$	& $277^{+186(435)}_{-289(681)}$	& $-0.00^{+-0.46(1.99)}_{-5.45(8.03)}$	& $0.02^{+0.03(2.90)}_{-3.35(5.69)}$	& -	\\[0.25cm] 
Tuc4	& $14^{+125(278)}_{-86(196)}$	& $-2.28^{+0.53(1.34)}_{-0.52(1.17)}$	& $-1.34^{+0.61(1.41)}_{-0.63(1.39)}$	& $250^{+116(269)}_{-182(559)}$	& $-0.00^{+-1.20(0.01)}_{-2.80(3.62)}$	& $0.02^{+-0.60(0.07)}_{-1.79(2.57)}$	& -	\\[0.25cm] 
Tuc5	& $17^{+112(257)}_{-82(183)}$	& $-2.05^{+0.38(0.94)}_{-0.44(0.96)}$	& $-0.96^{+0.51(1.20)}_{-0.57(1.25)}$	& $227^{+116(256)}_{-156(520)}$	& $-0.00^{+-1.14(0.00)}_{-2.40(3.06)}$	& $0.02^{+-0.43(0.11)}_{-1.44(2.05)}$	& -	\\[0.25cm] 
Gru2	& $-143^{+130(308)}_{-86(196)}$	& $-2.30^{+0.41(0.94)}_{-0.51(1.11)}$	& $-0.27^{+0.52(1.16)}_{-0.54(1.27)}$	& $38^{+182(377)}_{-172(458)}$	& $-0.00^{+-1.47(0.00)}_{-3.08(4.00)}$	& $0.02^{+-0.05(0.68)}_{-1.22(2.05)}$	& -	\\[0.25cm] 
Cet2	& $-171^{+206(451)}_{-154(333)}$	& $-2.82^{+1.04(2.52)}_{-1.26(2.85)}$	& $-2.38^{+1.01(2.26)}_{-1.18(2.57)}$	& $-53^{+239(516)}_{-201(473)}$	& $-0.00^{+-1.75(0.02)}_{-4.76(6.60)}$	& $0.02^{+-0.99(0.02)}_{-2.94(4.38)}$	& -	\\[0.25cm] 
Col1	& $265^{+91(186)}_{-88(203)}$	& $-0.21^{+0.17(0.35)}_{-0.12(0.27)}$	& $-0.50^{+0.25(0.47)}_{-0.11(0.29)}$	& $339^{+91(186)}_{-137(590)}$	& $-0.00^{+0.01(0.27)}_{-0.63(1.23)}$	& $-0.46^{+0.25(0.48)}_{-0.23(0.59)}$	& -	\\[0.25cm] 
Ind2	& $-79^{+95(208)}_{-63(139)}$	& $-0.53^{+0.09(0.27)}_{-0.11(0.26)}$	& $0.08^{+0.09(0.18)}_{-0.12(0.25)}$	& $-32^{+144(373)}_{-94(289)}$	& $-0.53^{+0.08(0.53)}_{-0.21(0.65)}$	& $0.02^{+0.16(0.29)}_{-0.15(0.80)}$	& -	\\[0.25cm] 
	\hline
	\end{tabular}
\end{table*}

\begin{figure}
\includegraphics[width=\columnwidth]{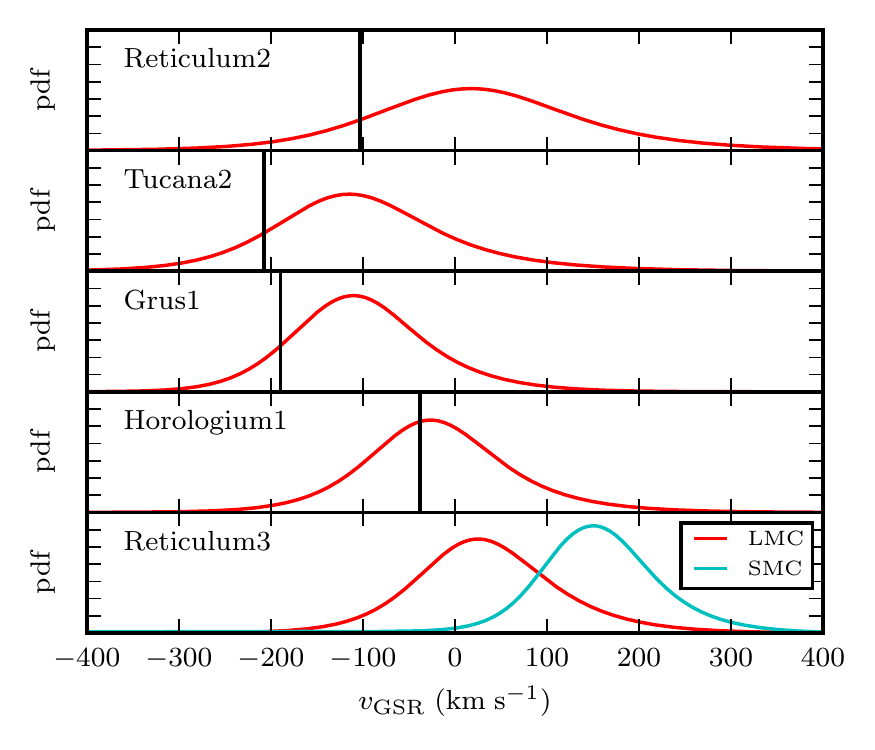}
\caption{Predictions for the GSR line-of-sight velocity for five of the DES satellites as labelled in each panel.
The y-axis ranges from 0 to 0.007 in each panel.
The red/cyan distributions are predictions assuming membership to the LMC/SMC populations.
The black vertical lines show measured values.
}
\label{fig:velocities_compare}
\end{figure}

Table~\ref{tab:velocities} shows our resulting predictions the for line-of-sight velocity and two components of proper motions in the solar rest-frame.
We report the maximum likelihood values along with the 68\% and 95\% confidence intervals.
The rightmost column shows measured values where available.
Fig~\ref{fig:velocities_compare} shows the predictions assuming LMC membership, transformed into GSR, for the four satellites with measured line-of-sight velocities.
The breadth of these distributions is due to the combination of the inherent velocity dispersion in our MC haloes and the spread in possible MC orbital histories.
We see that all four measured velocities are consistent with being drawn from the LMC population at the $2\sigma$ level.
In particular, Horologium 1 lies at the peak of its LMC satellite velocity prediction, taking its probability of LMC membership beyond reasonable doubt.

The bottom panel of Fig.~\ref{fig:velocities_compare} shows the predictions for Reticulum 3, the one satellite with a reasonable chance of belonging to the SMC population based on its position, according to Fig.~\ref{fig:satprob_r}.
Though the velocity distributions assuming LMC and SMC membership are somewhat coincident, a measurement of $v_\mathrm{GSR} > 100$ km s$^{-1}$ (i.e. $v_\mathrm{hel} > 270$ km s$^{-1}$) for Reticulum 3 would increase the likelihood of its belonging to the SMC.
As will discuss in Section \ref{sec:conclusions}, proper motions would also help us understand their origin.

\section{The Mass of the LMC}
\label{sec:lmcmass}

While the enclosed mass of the LMC is well constrained at $M(<8.7\;\mathrm{kpc})=(1.7\pm0.7)\times10^{10}M_\mathrm{\sun}$ \citep{vdmarel14}, its total mass remains largely unknown.
Abundance-matching would suggest that a galaxy with the stellar mass of $\sim2.9\times10^9M_\mathrm{\sun}$, equal to that of the LMC's \citep{vdmarel02}, are likely to reside in haloes with a virial mass of $2\times10^{11}M_\mathrm{\sun}$ \citep{moster13}.
Evidence for such a mass, however, is scarce.
Circumstantial evidence comes from models of the Magellanic Stream formation which require the LMC and SMC to form a long-lived binary pair \citep{besla12}, which is the favoured scenario for $>10^{11}M_\mathrm{\sun}$ LMCs \citepalias{kallivayalil13}.
\citet{penarrubia15} attempted a direct measurement by modelling the LMC's dynamical effect on galaxies in the Local Volume.
They inferred a mass $2.5\pm0.9\times10^{11}M_\mathrm{\sun}$, but stressed this result must be treated with caution due to the simplifying assumptions necessary in their analysis.

The LMC's satellite galaxies, acting as tracers of its gravitational potential at large distances, offer a new line of inquiry.
Having demonstrated, in Section~\ref{sec:3dmodel}, that many of the DES satellites have a high probability of belonging to such a population of LMC satellites, we will now try to use them to infer the LMC mass.

\subsection{$M_\mathrm{LMC}$ estimate based on positions only}

We first ask what constraints we can put on the LMC mass before assuming membership or otherwise of any satellite to the LMC population.
We calculate the posterior pdf of $M_\mathrm{LMC}$ from our three component (MW+LMC+SMC) 3D spatial distribution model
\begin{equation}, 
P(M_\mathrm{LMC}|d_\mathrm{DES}) =  \frac{1}{(*)}\int\limits_{\boldsymbol\Theta} \mathcal{L}(d_\mathrm{DES}|\boldsymbol\Theta) P(M_\mathrm{LMC}|\boldsymbol\Theta) P(\boldsymbol\Theta) \mathrm{d}\boldsymbol\Theta, 
\end{equation}
where $(*)$ is the denominator of Equation~(\ref{eqn:bayes}).

\begin{figure}
\includegraphics[width=\columnwidth]{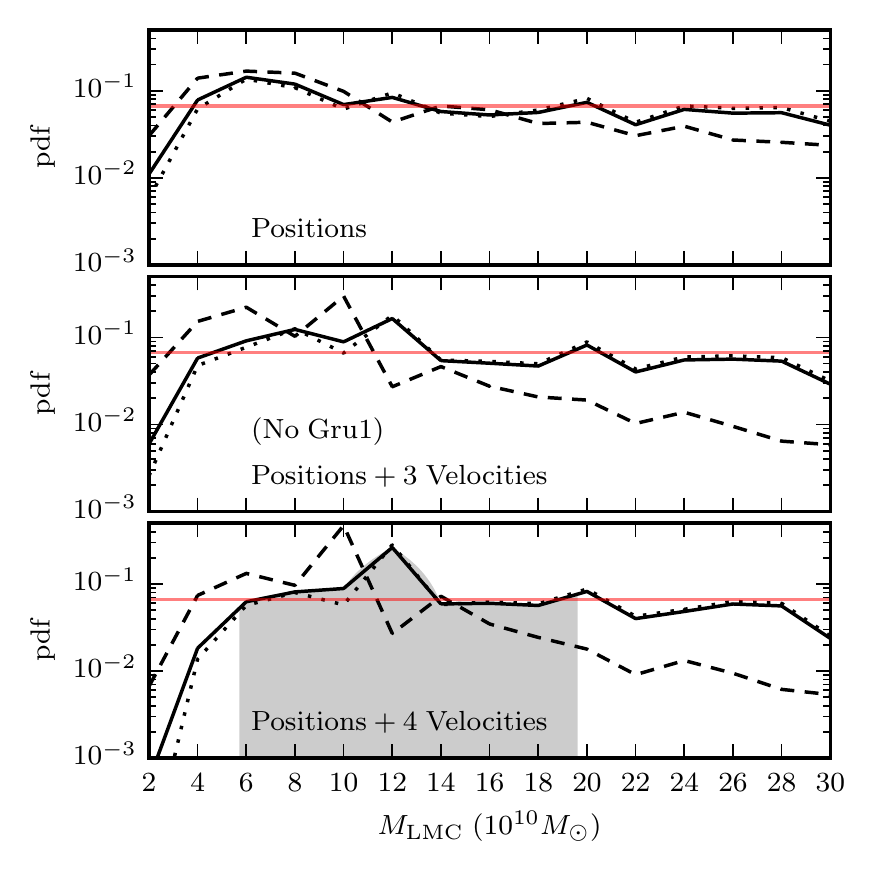}
\caption{Posterior pdfs on the virial mass of the LMC halo.  The top row
  uses just the positions of the DES satellites, the middle/bottom rows
  include velocity information from 3/4 satellites.  Dashed/dotted
  black lines show the results using cored/cuspy initial conditions,
  solid lines show the marginalised result, red line shows the flat
  prior.  The shaded region in the bottom panel shows the 68\%
  confidence interval about the most likely value.  }
\label{fig:mlmc}
\end{figure}

The top panel of Fig.~\ref{fig:mlmc} shows the result.
The posterior is broadly flat, which may be expected when attempting to constrain a galaxy's mass without the use of velocity information.
There is, however, a small peak between $4\leq M_\mathrm{LMC}/10^{10}M_\mathrm{\sun}\leq 6$ whose origin we briefly discuss.

The constraining power at the low mass end is the combined result of two effects.
Firstly, our lowest mass LMC model is likely to play host to the fewest number of satellites, which we impose via the prior shown in Equation~(\ref{eqn:newprior}).
Secondly, the same LMC is also the most likely to undergo multiple pericentric passages with the MW, seen in Fig.~\ref{fig:nperi}, as a result of the mass dependence of dynamical friction.
The likely fate for the relatively few satellites of a $2\times 10^{10} M_\mathrm{\sun}$ LMC, then, is to be tidally stripped from their host during a previous close approach with the MW, in slight tension with the observed concentration of satellites around the LMC.

The constraining power at the high mass end comes from the extent of the observed satellite distribution.
As can be seen in Fig.~\ref{fig:satprob_r}, 12 out of 14 DES satellites are within 100 kpc of the LMC, while 9 are within 50 kpc.
Given that our simulated satellite distributions extend out to the virial radius of their host, we slightly disfavour very massive LMCs which have virial radii extending far beyond the observed satellite distribution.

\subsection{$M_\mathrm{LMC}$ estimate using velocity information}

We now make the assumption that some of the DES satellites are indeed satellites of the LMC, and use their measured velocities to constrain its mass.
Section~\ref{sssec:magprob} shows that based on their 3D positions alone, of the four satellites with measured velocities, three are likely members of the LMC while one, Grus 1, is a possible member.
We therefore perform this analysis both with and without Grus 1.
It is important to note that our assumption here is that the satellites have existed in the LMC halo at some point, not the stronger statement that they are still bound to the LMC today.
Our simulations explicitly include satellites which have been, or are currently being, tidally stripped from the LMC by the MW.

To incorporate the velocity information we define a new likelihood function.
Of the 14 satellites whose positions in Section~\ref{sec:3dmodel}, we have now assumed that $N$ of these, which have measured velocities, are LMC satellites.
Our new, combined data set thus comprises
\begin{enumerate}
\item$d_L = \{\mathbf{x}_L,M_{V,L},v_L\}$: the positions, luminosities and velocities of $N$ satellites assumed to belong to the LMC
\item$d = \{ \mathbf{x}, M_V\}$: the positions and luminosities of the remaining DES satellites
\end{enumerate}
The new likelihood function is given by
\begin{align}
\log \mathcal{L}(d,d_L&|\boldsymbol\Theta) = \mathrm{Equation~(\ref{eqn:likelihood_pos})} \; \mathrm{summed} \; \mathrm{over} \; d \; + \\
	& \sum_{i=1}^{N} \log F_L(\mathbf{x}_L^i) + \sum_{i=1}^{N} \log P(v_L^i|\mathbf{x}_L^i,\boldsymbol\Theta,\mathrm{LMC}),  \nonumber
\end{align}
where Equation~(\ref{eqn:likelihood_pos}) refers to the original likelihood using just positional data, $F_L(\mathbf{x})$ is LMC component of our satellite distribution model given by Equation~(\ref{eqn:totmod}), and  $P(v|\mathbf{x},\boldsymbol\Theta,\mathrm{LMC})$ are the the local velocity distributions approximated from the simulations as described in Section~\ref{sec:veloc}.
We calculate posteriors on $M_\mathrm{LMC}$ using this new likelihood.

The bottom two rows of Fig.~\ref{fig:mlmc} show the result when we include velocity information.
In each panel the dashed and dotted lines show the results when we use the cored or cuspy satellite populations respectively, while the solid line shows the marginalised result.
As in Fig.~\ref{fig:N_posterior}, the mean posterior closely follows the results using the cusped profile since these typically provide better fits to the data.
Despite the variation between results dependent on our choice of initial satellite density profile, and a level of noise seeded from noise in our simulations, we find that some trend persists.

The addition of the 3 velocities of likely LMC satellites leads us to slightly favour masses $M_\mathrm{LMC}<14 \times 10^{10} M_\mathrm{\sun}$.
This weak upper bound is mainly driven by Horologium 1 which, as discussed in \citet{koposov15b}, has a line-of-sight velocity just 15 km s$^{-1}$ discrepant from the value expected at its position if it were stationary with respect to the LMC.
Such a low relative velocity is more likely to occur in lighter LMCs, which have smaller halo velocity dispersions.
The addition of Grus 1 to our velocity sample creates a peak in the posteriors in the range $10 \leq M_\mathrm{LMC}/10^{10} M_\mathrm{\sun} \leq 12$, and leads us to strongly disfavour the lightest LMC, $M_\mathrm{LMC}=2 \times 10^{10} M_\mathrm{\sun}$, which we discuss below.
Assuming all four satellites do indeed belong to the LMC, the most likely value and 68\% confidence interval for the LMC halo mass is given by $M_\mathrm{LMC} = 12^{+8}_{-7} \times 10^{10} M_\mathrm{\sun}$.

\subsection{Grus 1}

\begin{figure*}
\includegraphics[width=\textwidth]{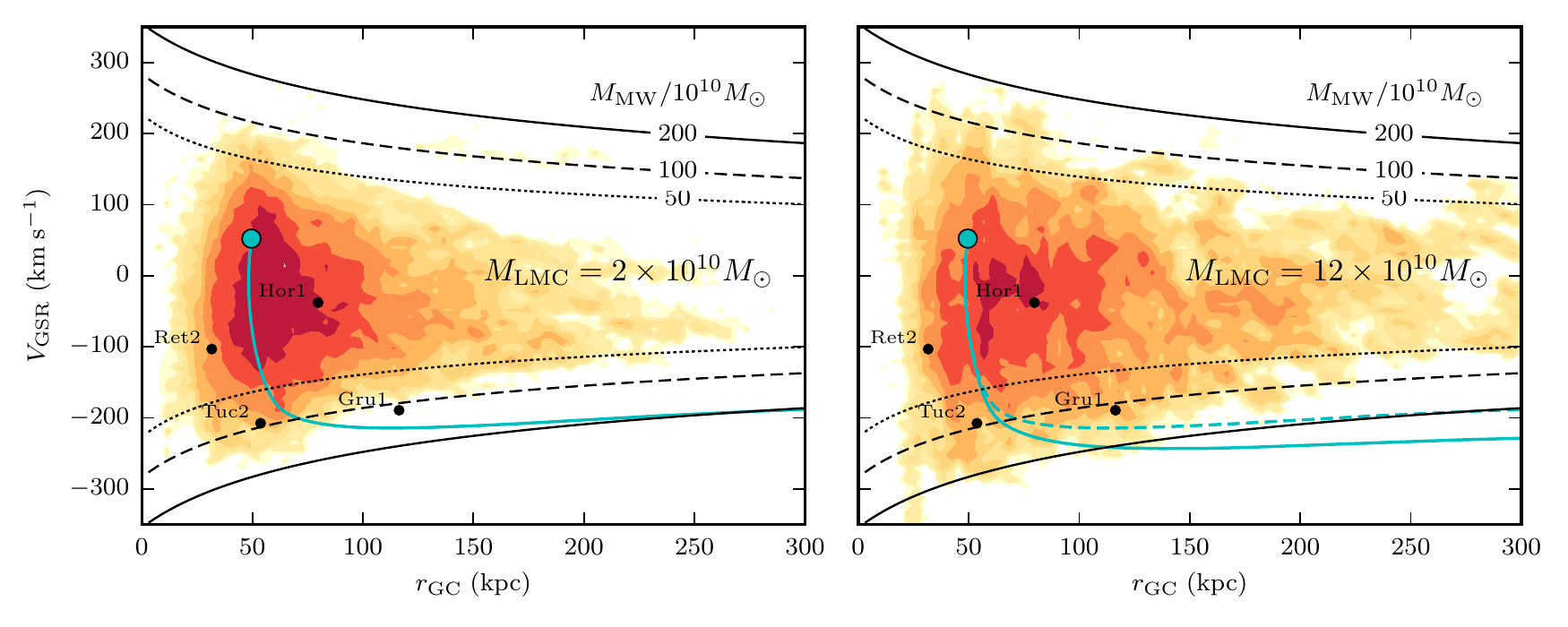}
\caption{
Phase space diagram showing line of sight velocity in GSR against Galactocentric distance.
Filled contours shows the simulated satellite distributions for a $2\times 10^{10} M_\mathrm{\sun}$ (left panel) and $12\times 10^{10} M_\mathrm{\sun}$ (right panel) LMC.
Also shown are the positions of the four DES satellites with measured velocities (black dots) and escape velocity curves for three NFW haloes, with masses labelled and concentrations given by Equation~(\ref{eqn:cM}), scaled by a factor $1/\sqrt{3}$ to account for unknown tangential velocity components.
The LMC is shown by a cyan circle, its past orbit as calculated for the mass given in each panel is shown by a solid cyan line.
The orbit in the left panel is repeated as a dashed line in the right panel for comparison.
}
\label{fig:rv}
\end{figure*}

Figure~\ref{fig:rv} demonstrates how the inclusion of Grus 1 imposes a lower bound on the LMC mass.
We take simulations of a light and heavy LMC, with masses of 2 and 12 $\times 10^{10} M_\mathrm{\sun}$ respectively, with all other model parameters kept fixed.
For particles inside the DES footprint, we plot the phase space distributions in Galactocentric $r$ against $v_\mathrm{GSR}$, and over-plot the phase space positions of the DES satellites with measured velocities.
At the distance of Grus 1, the satellite distribution for the light LMC does not extend to sufficiently negative velocities to accommodate Grus 1, yet it is well within the scope of the heavy LMC's distribution.

There are two reasons for this.
Firstly, the satellites of the heavy LMC comprise a hotter population than those of the light LMC, filling a larger volume of the available phase space. 
The second reason is due to the mass dependence of the LMC's orbital history, demonstrated in the right panel of Fig.~\ref{fig:rv} where we compare the past orbits of the light LMC (dashed line) with that of the heavy LMC (solid line).
Both are calculated in the potential of our Light MW model for $z=0$ velocities given by the central values in Table~\ref{tab:kinematics}.
Integrating backwards from this fixed velocity, dynamical friction accelerates the LMC along its orbit.
The heavy LMC, experiencing more dynamical friction than the light, thus has a larger (i.e. more negative) velocity than the light LMC at the same radius.
This is reflected in the mean velocities of their satellite distributions.

We now re-consider Grus 1's inclusion in our velocity sample.
Figure~\ref{fig:satprob_r} suggests that based on its position, Grus 1 is almost as likely to be a MW as an LMC satellite.
We draw some reassurance that the latter is true from Grus 1's velocity, which appears in Fig.~\ref{fig:rv} to lie above the escape velocity of a $M_{200}=10^{12}M_\mathrm{\sun}$ NFW halo.
\citet{boylankolchin13} (henceforth BK13) showed that it is vanishingly rare for a host halo to contain an unbound subhalo within $\sim1.5$ virial radii.
Assuming a virial mass no greater than than $10^{12}M_\mathrm{\sun}$ for the MW - as has been inferred using stellar tracers \citep[e.g.][]{williams15}, tidal streams \citep[e.g.][]{gibbons14} and Local Group kinematics \citep[e.g.][]{diaz14} - this implies that Grus 1 is very unlikely to belong to the MW population.
Instead, the apparent ``unboundedness'' of Grus 1 is likely the result of its recent accretion in a massive group.
At $z=0$, such an event is underway in one out of six simulated haloes in BK13; the LMC serves as an analogous host to Grus 1 in the MW.

On a similar note, in order to explain the high 3D velocity of Leo I \citep{sohn13}, BK13
proposed the alternative hypothesis that the MW halo is significantly
more massive than $10^{12}M_\mathrm{\sun}$.  It may be necessary to
re-visit this conclusion in light of our inference from
Section~\ref{sssec:nmagsats} that $\sim45$ satellites have been
accreted with the MCs, i.e. there likely exists a significant
contamination to the MW satellite population.

\subsection{Meaning of the Derived Mass}

The mass constraints we have derived for the LMC (see Fig. \ref{fig:mlmc}) raise the question, how does this mass correspond to the LMCs mass today?
The LMC mass we use in our simulation affects the satellite distribution today in two ways.

Firstly, the LMC mass affects the dynamics of the MW and LMC (and SMC).
As we discussed in Section \ref{ssec:galaxymodels}, we use a constant mass for the MW and LMC (and SMC) during their 10 Gyr evolution.
While their real masses are evolving over these timescales, if the mass which will eventually accrete onto the MW remains within the LMC's orbit and is reasonably isotropic for most of this time, we can think of the MW as having a constant mass.
Such reasoning also explains why the Timing Argument, which models the MW and M31 as un-evolving point-masses for the age of the Universe, provides a good estimate for the mass of the Local Group \citep{li08}.
Such an argument also holds for the LMC.
Thus, the mass inferred from the LMC's dynamics should be similar to the LMC's total mass before it began disrupting in the MW.

Secondly, the LMC's mass affects the number of LMC satellites, the radial distribution of these satellites, and their subsequent dynamical evolution.
The number of satellites assigned to the LMC is guided by the maximum number of satellites the LMC had which is controlled by the LMC mass before it began disrupting in the MW.
Similarly, the radial profile of the satellites around the LMC is controlled by properties of the LMC before infall onto the MW.
The effect of the LMC's mass on the satellite's dynamics is difficult to gauge since much of the satellite stripping can occur within the MW potential, when the LMC has been substantially stripped.
Thus, we expect the LMC mass constraints derived here to correspond to the LMC's mass shortly before it began disrupting in the MW. We will compare the mass estimate here against cosmological zoom-in situations in future work.

\section{Discussion and Conclusions}
\label{sec:conclusions}

We have built a dynamical model of the satellite populations of the
MCs based on their observed kinematics and incorporating
assumptions based on the results of cosmological simulations where
required. Specifically, our model includes the response of the MW to
the in-falling MCs as well as the effects of the dynamical friction
acting on the Clouds. Given that the initial intrinsic density
distribution of the satellites in each of the Clouds is unknown, we
have allowed for two distinct radial profiles. Finally, we have
emulated the effects of tidal shredding by introducing a destruction
region within each Cloud. Our numerical simulation suite samples over
the proper motion values allowed by the recent HST observations of the
MCs and runs across a range of masses for each of the 3 galaxies (the
MW, the LMC and the SMC). We compared this model against the
satellites discovered in the DES survey. Here we summarise our
results, after discussing the limitations of our model.

\subsection{Shortcomings}
\label{ssec:shortcomings}

First and foremost, our simulations lack any treatment of the mass evolution of the MW, LMC or SMC over their 11.5 Gyr timespan.
A self-consistent treatment of this effect requires the use of zoom-in high-resolution N-body simulations, but given our wish to prioritise observed MC kinematics, coupled with the dearth of LMC analogues in simulated MW-like haloes \citep{boylankolchin13}, this precludes the use of such simulations here.

An alternative approach to model the MW mass evolution is to vary the parameters of analytic potentials \citep[e.g. as in][]{yozin15}.
We have chosen against doing this because the majority of our parameter space has the LMC on its first passage about the MW and for these cases, the mass which the MW will accrete prior to the arrival of the LMC will at all times have been contained within the LMC's orbit.
Assuming, for lack of better constraints, that this mass is isotropically distributed about the MW we may as well assume it has been concentrated at the MW for the entirety of the time considered.
Such reasoning also explains why the point mass approximation provides a good estimate to the mass of the Local Group when employed in the Timing Argument \citep{li08}.

For solutions with multiple LMC pericenters, \citetalias{kallivayalil13} find that including MW mass growth increases the LMC's orbital period by $\sim30\%$ and eccentricity by $\sim10\%$.
From our N-body experiments described in Section~\ref{ssec:dynfric}, we find similar deviations arise due to the force of tidally stripped LMC material on its bound core.
Though simple prescriptions exist for modelling tidal mass loss \citep[e.g.][]{zentner03} and the role of tidally stripped material on subsequent orbital evolution \citep[e.g.][]{fellhauer07}, their efficacy for massive, LMC-like halos is questionable.
Pending the development of more robust models at these mass scales, the trustworthiness of our calculated orbits beyond a few Gyr remains in doubt.

Having said this, since our MC models are grounded in their observed kinematics, revisions to their orbital history will least affect the distribution of satellites which have been stripped most recently from - or indeed are still bound to - the MCs.
On average, these are the satellites nearest by.
Given the proximity of the DES satellites to the LMC - with 9 out of 14 within 50 kpc - unless explicitly stated below, we believe that our results should be robust to such revisions of the MC orbital histories.

Another limitation of our simulations is that, by representing satellites as tracer particles, they do not allow for satellite-satellite interactions.
In the mass regime of ultra-faint dwarfs, $\sim3\%$ of dwarf satellites in MW-like halos have undergone a major merger since $z=1$ \citep{deason14}, which we consider this a small effect justifiably omitted from our model.
We note however, the incidence of satellite mergers increases for more massive satellites \citep{wetzel15}. 

Lastly, our model assumes that the radial distribution of MW dwarfs follow the radial distribution of subhaloes in $\Lambda$CDM.
This assumption is largely untested.
The bottom left panel of Fig.~\ref{fig:mwmod} shows that dwarfs in SDSS-DR5 (grey histogram) appears more centrally concentrated than the $\Lambda$CDM prediction, even when accounting for observational limits (black dashed line).
Although the addition of baryons to $\Lambda$CDM suggests a flatter, rather than more concentrated, distribution of subhaloes \citep{donghia10}, we investigate how our conclusions change when we re-run our analysis using a more concentrated distribution of MW satellites (Einasto with parameters $\alpha=0.4$ and $r_{-2}=50$ kpc, the result of a maximum-likelihood fit to the SDSS-DR5 dwarfs in Fig.~\ref{fig:mwmod}).
Our prediction for the probability of an LMC origin for any of the DES satellites drops significantly: our 4 best LMC candidates have membership probabilities $0.3<p_L<0.6$, as opposed to 7 with $p_L>0.7$ in the earlier analysis.
Interestingly, however, our posterior on the total number of Magellanic satellites retains a peak at $\sim$70.
In short, models with a Magellanic contribution are still preferred to explain the DES satellites.
To break the degeneracy between the relative sizes of the MW and LMC contributions will likely require simultaneous consideration of the dwarf population over the whole sky.

\subsection{Principal results}

Our main conclusions are summarised below (errors represent 68\% confidence intervals, unless stated otherwise):
\begin{enumerate}
\renewcommand{\theenumi}{(\arabic{enumi})}
\item The MW dwarf galaxy population, if assumed to follow the $\Lambda$CDM radial subhalo distribution, is unable to reproduce the satellite counts observed in DES.
Our 68\% (95\%) confidence limits on the contribution of such a Galactic component to the number of satellites observed in DES is fewer than 4 (8) out of the 14 considered.
The posterior probability distribution for the MW contribution is shown in the left column of Figure~\ref{fig:N_posterior}.

\item The MCs have been accreted onto the MW with $70^{+30}_{-40}$
  satellite galaxies (middle column of Figure~\ref{fig:N_posterior}),
  many of which should reside in the leading arm of debris stretching
  behind the Galactic plane. Figure~\ref{fig:leadingarm} shows exactly where on the sky one should search to maximize the chances of uncovering
  this missing population.

\item Based on their 3D positions, of the 14 dwarf galaxy candidates
  discovered in the DES survey, 12 are possible ($p>0.5$) and 7 are
  likely ($p>0.7$) members of the LMC satellite population (see
  Figures~\ref{fig:satprob} and \ref{fig:satprob_r}, as well as
  Table~\ref{tab:plmc}). For the four satellites where data is available,
  all have velocities consistent with being LMC satellites (see
  Figure~\ref{fig:velocities_compare}).

\item The influence of the SMC on the properties of the Magellanic satellite
  population is minor in our numerical experiment. As a result the
  total number of the SMC satellites is unconstrained (see right panel
  of Figure~\ref{fig:N_posterior}). There is, however, a weak
  indication that Reticulum 3 could be considered as a possible
  candidate ($p=0.3$) to be an SMC satellite. A measurement of
  $v_\mathrm{hel}>100\;\mathrm{km}\;\mathrm{s}^{-1}$ would strengthen
  this case (see Fig. \ref{fig:velocities_compare}).

\item The newly discovered satellites do not uniformly cover the
  entire DES footprint: this strong apparent concentration of dwarfs
  around the LMC strengthens the case ($\Delta p=+0.26$, $p=0.90$)
  that the LMC has completed exactly one pericentric passage with the
  MW (see Table~\ref{tab:nperi}).

\item Seven of the DES satellites considered have
  correlated on-sky positions. Surprisingly, these lie in a 3 kpc
  thick, 90 kpc wide plane passing through the LMC.  If we assume that
  the satellites are drawn from a population initially distributed
  isotropically about the LMC, then evolved to today in the potentials
  of the MW, LMC and SMC, the probability that this is a chance occurrence
  is $p=0.05$. The SMC lies only 2 kpc from this plane.

\item Assuming that the four satellites with measured velocities are
  associated with the LMC (for which $p_\mathrm{prior}>0.7$ in three
  cases and $p_\mathrm{prior}=0.6$ for Grus 1) we can derive a
  constraint on the mass of the LMC halo of $M_{200} = 12^{+8}_{-7}
  \times 10^{10} M_\mathrm{\sun}$.  The necessary assumption here is
  that the satellites have evolved within the LMC halo prior to their
  arrival at the MW, not that they are still bound to the LMC today.
  The lower bound rests on Grus 1 taking part in the Magellanic
  in-fall.  Furthermore, given Grus 1's relatively large distance of
  $\sim$90 kpc from the LMC, this constraint is the one most sensitive
  to possible revisions to the orbital histories of the MCs.

\end{enumerate}

\subsection{Discussion}

To verify the Magellanic origin hypothesis, accurate 3D kinematics of
the newly detected satellites ought to be collected. As we show above,
radial velocities are discerning as the debris follow a particular
track in phase-space. However, given that most of the dwarfs are
at distances of order 50 kpc or further, the information as to the
exact direction in which they are traveling is encoded in the satellites'
proper motions. It is therefore the astrometric efforts, both current
utilising the HST \citep[see e.g.][]{kallivayalil13,sohn13,bellini14}
as well as the forthcoming ones taking advantage of Gaia and the LSST, 
that are of utmost importance. To facilitate the hypothesis testing, we
list the predicted satellites' line-of-sight velocities and proper
motion values in Table~\ref{tab:velocities}.

While it is plausible that the LMC contributed significantly to the
tally of the Galactic satellites by providing $\sim$70 low-luminosity
galaxies within the MW virial radius, it is unlikely to be responsible
for the overall anisotropy of the dwarf distribution around us. As
clearly illustrated in Figures~\ref{fig:mwmod} and
\ref{fig:N_posterior}, the SDSS-DR5 region of the sky appears to be
overdense in satellites compared to the SDSS-DR10 area increment, as
well as the PS3$\pi$ and DES footprints. We can calculate the number
of the SDSS-DR5 objects falling within the portion of the sky
enclosing the bulk the Magellanic debris in our simulations. The pole
of the best fitting debris plane is given by $(l,
b)=(353^{\circ},-12^{\circ})$ (see also
Figure~\ref{fig:leadingarm}). Within $\pm20^{\circ}$ of this plane,
however, there are only 2 faint (i.e. not classical) satellites - Leo
IV and Leo V. Moreover, the direction of motion of this pair could be
incompatible with that of the MCs, as recently speculated by
\citet{torrealba16}. Accordingly, as far as the ultra-faint dwarf
galaxies are concerned, their on-sky distribution appears to be much
broader than the tidal tails associated with the LMC accretion. The
asymmetric shape of the distribution is perhaps more consistent with
the (ubiquitous in $\Lambda$CDM) filamentary accretion rather than a
recent group in-fall scenario.

Nonetheless, the anisotropy of the Galactic satellite distribution
highlighted in this work could affect the estimate of the total number
of the MW satellites within its virial radius. For example, assuming
the satellite density normalization stipulated by the number of
satellite detections within the SDSS-DR10, in combination with the
detection threshold given by Eq.~\ref{eqn:dr5} and the radial
distribution given by Einasto profile as described in
Section~\ref{ssec:mwmod}, results in $\sim$200 satellites within 300
kpc (compared to $\sim$300 using the SDSS-DR5 density normalization,
see Figure~\ref{fig:mwmod}). Thus the ratio of the total number of
dwarf satellites could be as high as
$N_{\mathrm{LMC}}/N_{\mathrm{MW}}\sim 1/3$ while the host mass ratio
is only $M_{\mathrm{LMC}}/M_{\mathrm{MW}}\sim 1/10$. This estimate
comes with a large uncertainty, and as such should be taken with great
caution. If reliably established, the scaling of the number of
satellites with the host mass can offer a glimpse into the physics of
dwarf evolution in group environments, where the effects of
pre-processing have so far been largely neglected and are currently
poorly constrained \citep[see][]{wetzel15}. For example, it is
speculated that the low gas temperature and the remote Local Volume
location of the group halo could enhance the survival chances of the
low-luminosity dwarfs \citep[e.g.][]{donghia08}. Additionally, in
Magellanic analogs - compared to the MW analogs with masses of
$\sim10^{12}$M$_{\odot}$ - tidal stripping must be suppressed, as only
the very central parts of the host contain densities high enough to
start shredding the orbiting subhaloes. This can be contrasted with
the recent detection of an extended and highly substructured stellar
halo around the MCs by \citet{bk16}, naively implying that ample
stripping has been taking place in the LMC.

Inspired by the results presented here, the prospects for the future
study are plenty. We envisage a wide imaging campaign to pick up the
remaining LMC dwarfs, complemented by a comprehensive spectroscopic
follow-up to constrain their orbital properties. In parallel, the
effects of the (lack of) group pre-processing should be pinned down
with the help of high-resolution collisionless and hydrodynamical
simulations. Finally, the analysis described here should be extended to
the remainder of the Milky Way's satellite population.

\section*{acknowledgements}
The authors wish to thank the members of the Cambridge STREAMS club, in
particular Sergey Koposov, as well as Alex Drlica-Wagner for their insightful comments
that helped to improve this manuscript.
PJ thanks the Science and Technology Facilities Council (STFC) for the award of a
studentship. The research leading to these results has received
funding from the European Research Council under the European Union's
Seventh Framework Programme (FP/2007-2013) / ERC Grant Agreement
no. 308024.





\footnotesize{ \bibliographystyle{mn2e} \bibliography{mybib} }






\bsp	
\label{lastpage}
\end{document}